\documentclass[11pt]{article}

\usepackage[final]{acl}

\usepackage{times}
\usepackage{latexsym}
\usepackage{algorithm}
\usepackage{algorithmic}
\usepackage{amsmath}
\usepackage{amssymb}
\usepackage{enumitem}
\usepackage{multirow}
\usepackage{booktabs}
\usepackage{float}

\usepackage[T1]{fontenc}

\usepackage[utf8]{inputenc}

\usepackage{microtype}

\usepackage{inconsolata}

\usepackage{graphicx}

\title{Subtraction Gets You More: Gap-Aware Retrieval for Multimodal Multi-Hop QA}

\author{Sunah O \qquad Jay-Yoon Lee 
    \thanks{Correspondence author}
    \\
    Seoul National University 
    \\ \texttt{\{ec\_osa16586, lee.jayyoon\}@snu.ac.kr} 
 }

\begin{document}
\maketitle
\begin{abstract}
In multimodal multi-hop question answering, we focus on the initial retrieval stage via two distinct tasks: (1) evidence set completion, retrieving missing evidence given context, and (2) sequential pool construction, iteratively building the top-$K$ pool from the scratch. Under these settings, we point out that conventional iterative retrieval frameworks often suffer from Semantic Anchoring, where previously fetched evidence traps the retriever and yields entity-centric redundancy. To break this trap, we propose \textbf{GRAIL} (\textbf{G}ap-aware \textbf{R}etrieval via \textbf{A}daptive \textbf{I}mplicit \textbf{L}ocalization), a paradigm that performs implicit query rewriting directly at the embedding level. By context-subtractive query steering, GRAIL excels at compositional cross-modal reasoning, while additive embedding updates show strength on localized information aggregation.
To leverage these complementary strengths, we propose a Task-Adaptive Hybrid Framework, achieving a \textbf{35.7\%} macro-averaged performance gain on the MultimodalQA evidence set completion task.
Extensive evaluations demonstrate that sequential GRAIL retrieves in a superior, noise-resilient manner, significantly expanding the search horizon through iterative gap-aware optimization.
\end{abstract}

\section{Introduction}
\begin{figure*}[t]
  \includegraphics[width=\textwidth]{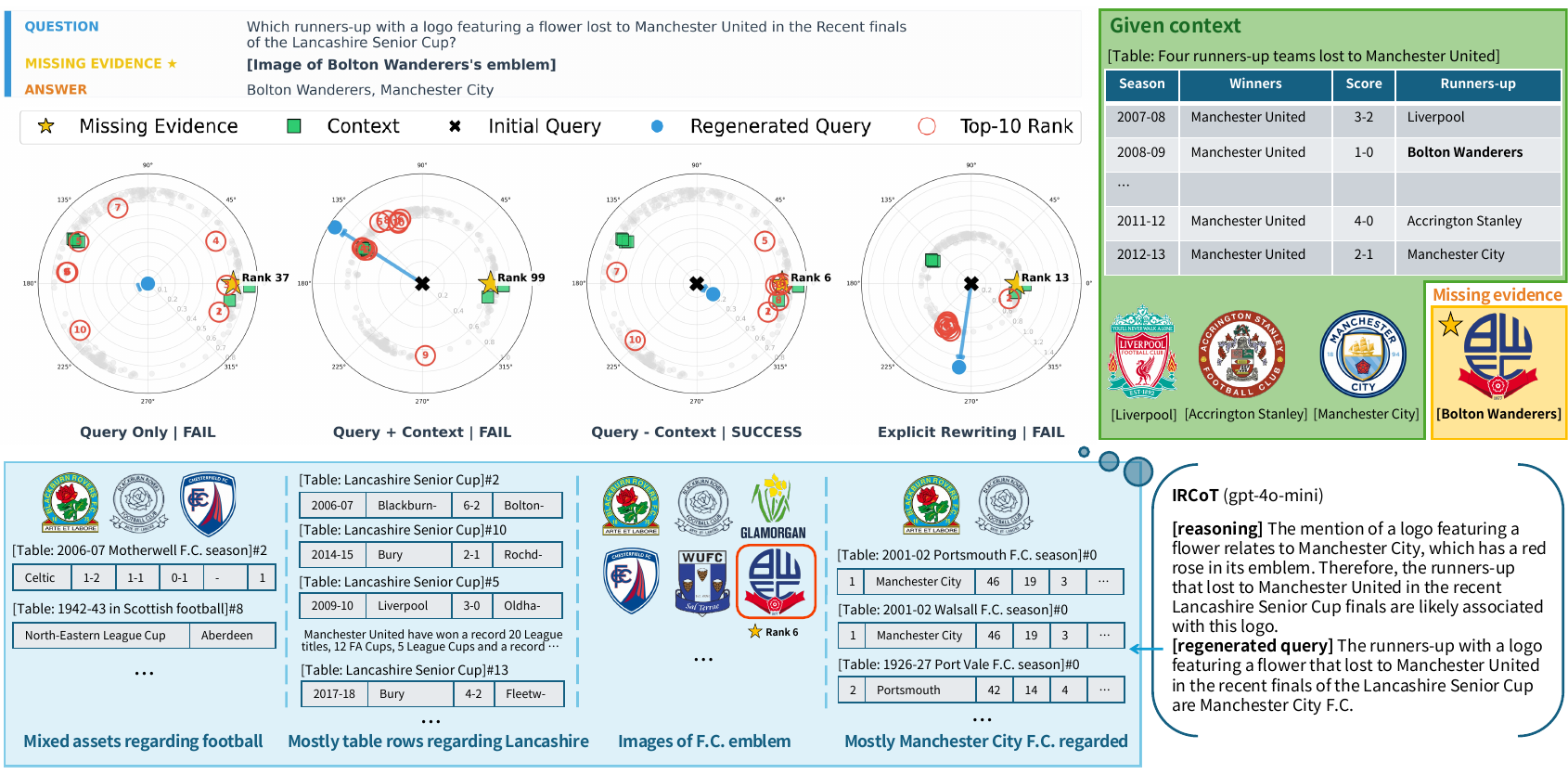}
  \caption{\textbf{Redundancy trap in MMQA.} Given a table and three emblems, the goal is to fetch the missing visual evidence (\textit{Bolton Wanderers's emblem}) to complete the evidence set. (i) Query $+$ Context fail as they trapped by the existing \textit{Lancashire Senior Cup} context (Semantic Anchoring). Meanwhile, Explicit Rewriting (IRCoT) prematurely concludes the search upon verifying \textit{Manchester City}'s record, overlooking the unexamined visual assets. (ii) Query $-$ Context succeeds by implicitly subtracting the covered context features within the embedding space, shifting the search direction toward the unobserved visual target. See Appendix~\ref{sec:app_case_study} for another case study.}
  \label{fig:intro}
\end{figure*}
In multimodal retrieval-augmented generation frameworks, the reasoning performance of Large Multimodal Models (LMMs) is fundamentally contingent on the quality of the evidence pool. While text-only tasks often allow models to rely on broad parametric knowledge, multimodal multi-hop questions answering (MMQA) demands a direct inspection of explicit, modality-specific details. Without having these context-dependent facts provided, LMMs cannot infer the high-density visual or structural details--such as specific trends in a chart or the color of objects--that are crucial for accurate reasoning. Consequently, retrieval is not merely a preliminary step; it effectively sets the performance ceiling of the entire retrieval-augmented LMM.

To address multi-step dependencies, a single, initial query is inherently insufficient to cover the entire sequence of required evidence. A straightforward alternative is to decompose the complex question a priori into independent sub-queries and execute them sequentially \cite{talmor2021multimodalqa}. Yet, such static decomposition frameworks immediately founder when confronted with open-domain multi-modality. In a vast multimodal corpus, it is nearly impossible to predict which modality--whether text, image, or table--contains the specific piece of evidence required for each hop. Because the modality of target facts remain unknown before the search begins, pre-determining a rigid textual decomposition pathway is structurally unviable.

To overcome this predictability barrier, a more dynamic paradigm has emerged: given an initial subset of evidence, subsequent queries are iteratively regenerated by adding the acquired context. However, as illustrated in Figure~\ref{fig:intro}, this evidence-additive query regeneration loop exhibits a critical structural failure we define as Semantic Anchoring. Instead of expanding the search space, the strong semantic gravity of previously fetched evidence traps the retriever, forcing an over-reliance on prominent entities (e.g., Lancashire Senior Cup). This leads to entity-centric redundancy, blindly fetching redundant assets while pushing the true target evidence--\textit{the complementary fact regarding Bolton's emblem}--out of the top-$K$ rank.

One might also ask why not simply rely on advanced LMMs for explicit query regeneration. In practice, forcing dense visual observations through a narrow textual detour introduces non-trivial token overhead and a high susceptibility to semantic drift. When an explicit text-generation agent prematurely commits to an incorrect textual path--such as anchoring heavily on \textit{Manchester City} (Figure~\ref{fig:intro})--the trajectory error compounds catastrophically. The language-centric bias misdirects subsequent search iterations toward redundant or irrelevant spaces, proving that explicit textification fails to capture unexamined, complementary multi-modal assets.

We introduce \textbf{GRAIL} (\textbf{G}ap-aware \textbf{R}etrieval via \textbf{A}daptive \textbf{I}mplicit \textbf{L}ocalization), a new paradigm for implicit query rewriting within the latent space. Rather than blindly contextualizing existing context, the rewritten query consistently shifts toward the unanswered facets of the question, increasing its similarity to missing evidence by subtracting the already covered information from the query. Operating entirely in the embedding space, this mechanism avoids any linguistic loss and enables an efficient, self-correcting retrieval trajectory for complex MMQA. When applied iteratively, it naturally supports progressive evidence discovery without accumulating semantic bias.

The primary contributions are as follows:
\begin{itemize} [leftmargin=*]
    \item \textbf{Problem Identification}: We systematically characterize Semantic Anchoring in MMQA, demonstrating how conventional iterative loops suffer from entity-centric redundancy (Figure~\ref{fig:intro}).
    \item \textbf{Implicit Latent Steering}: We propose \textbf{GRAIL} that performs context-subtractive query rewriting directly in the latent space, effectively bridging cross-modal reasoning gaps and mitigating semantic anchoring in additive retrieval.
    \item \textbf{Task-Adaptive Hybrid Framework}: Based on a comparative analysis of where gap-aware versus additive search excels, we design a hybrid framework that dynamically routes queries to handle both evidence aggregation and missing evidence retrieval. 
    \item \textbf{Robust Iterative Retrieval}: When applied iteratively on multimodal evidence, our mechanism constructs high-quality Top-$K$ candidate pools; unlike conventional additive methods prone to compounding errors, GRAIL remains robust to noise and prevents error propagation.
\end{itemize}

\section{Related Work}
\paragraph{Multimodal Question Answering} Early works in MMQA primarily focus on downstream reasoning or query decomposition under heavily constrained, "closed-world" environments. Frameworks such as AutoRouting \cite{talmor2021multimodalqa}, ImplicitDecomp \cite{talmor2021multimodalqa}, PReasM \cite{yoran-etal-2022-turning}, SKURG \cite{yang-2023-skurg}, and RAMQA \cite{bai-2025-ramqa} inherently rely on pre-filtered asset pools provided by the dataset metadata, without executing open-domain candidate retrieval. 
Even within these oracle-guided pools, PReasM \cite{yoran-etal-2022-turning} restricts its reasoning scope to text and table only, while RAMQA \cite{bai-2025-ramqa} is limited to single-hop QA. Furthermore, although MuRAG \cite{chen-2022-murag} scales to full-wiki image-text corpora, it lacks iterative reasoning across heterogeneous modalities.
While existing networks excel at localized reasoning under distractor settings, they do not explicitly address the challenges of open-domain, multimodal, multi-hop retrieval. More recently, frameworks like LILaC \cite{yun-2025-lilac} have introduced explicit LLM-based query decomposition coupled with structural graph traversal. However, since discrete planning systems are vulnerable to cascading planning errors, we execute implicit query steering directly within the embedding space.

\paragraph{Multi-hop Retrieval} Within the text-only domain, iterative retrieval and query reformulation--such as recursive vector addition in MDR \cite{xiong-2020-answering}, explicit entity expansion in GoldEn \cite{qi-2019-answering}, or LLM-based generation in IRCoT \cite{trivedi-etal-2023-interleaving} and IterDRAG \cite{yue2025iterdrag}--have widely pioneered sequential context construction.
However, translating text-centric paradigms into multimodal contexts introduces severe bottlenecks. 
Explicit textification of high-dimensional visual features is computationally expensive, flattens non-verbal cues into rigid keywords, and triggers compounding hallucinations or semantic drift.

\paragraph{Latent Space Purification} We introduce subtractive query steering, conceptually rooted in Rocchio relevance feedback \cite{rocchio-1971-relevance} and its dense retrieval extensions via contrastive learning and hard negative mining \cite{karpukhin-2020-dense, xiong-2020-approximate, izacard-2021-unsupervised}. 
While conventional multi-hop methods refine queries through the additive accumulation of retrieved evidence \cite{xiong-2020-answering}, our framework executes online, implicit subtraction to dynamically suppress redundant context and prevent semantic drift. 
Importantly, unlike classic Rocchio feedback, it eliminates explicit centroid updates or static weighting, learning to purify the multimodal latent space via a differentiable, step-wise gated transformation.

\section{Problem Formulation}
\subsection{Evidence Set Completion} \label{subsec:evidence_comp}
Multimodal multi-hop question answering (MMQA) requires integrating multiple pieces of evidence across diverse modalities to derive a concise answer. In an open-domain setting, this process necessitates retrieving a set of relevant items from a vast, multi-modal corpus $\mathcal{C}$. While previous studies often focus on the downstream reasoning stage, we shift our attention exclusively to the retrieval phase: the fundamental process of gathering the necessary information to enable such multi-hop reasoning.

Let $q$ be a query representation and $\mathcal{E}^* = \{e_1, e_2, \dots, e_G\}$ be the gold evidence set required to resolve $q$. Assuming a partial evidence subset $\mathcal{E}_p \subset \mathcal{E}^*$ has already been acquired, the retriever's objective is to identify the missing evidence piece $e_t \in \mathcal{E}^* \setminus \mathcal{E}_p$ from $\mathcal{C}$. Unlike standard retrieval paradigms that optimize for individual asset relevance in isolation, evidence set completion demands a complementary optimization of the candidate pool. The primary goal is to learn a mapping function $f(q, \mathcal{E}_p) \to q'$ that generates a gap-aware, reformulated query, which adaptively suppresses the information components already covered by $\mathcal{E}_p$.

\subsection{Sequential Pool Construction} \label{subsec:pool_const}
A practical open-domain MMQA scenario does not provide any partial gold context a priori. Therefore, we extend the single-step completion into a sequential retrieval task, where the retriever must autonomously construct an evidence set $\mathcal{E}_{\text{ret}}$ ($|\mathcal{E}_{\text{ret}}| = K$) starting from a query-only state ($\mathcal{E}_p = \emptyset$).

At the initial hop $t = 0$, the retriever dispatches $q$ to fetch an initial chunk of base evidence $\mathcal{E}_{\text{base}}$ of size $A$ ($< K$) from the corpus $\mathcal{C}$. The partial evidence set is initialized with this base context, establishing $\mathcal{E}_p^{(0)} = \mathcal{E}_{\text{base}}$. At each subsequent hop $t \in \{1, 2, \dots, T\}$, the mapping function processes the original query along with the accumulated context to formulate an updated retrieval vector:
\begin{equation}
    q^{(t)} = f(q, \mathcal{E}_p^{(t-1)}).
\end{equation}
This vector is deployed to retrieve a complementary block of evidence $\mathcal{B}^{(t)}$ of size $M^{(t)}$. The acquired context is then dynamically expanded via set union:
\begin{equation}
    \mathcal{E}_p^{(t)} = \mathcal{E}_p^{(t-1)} \cup  \mathcal{B}^{(t)}.
\end{equation}
The iterative process terminates when the total accumulated evidence fills the remaining slots ($\sum_{t=1}^{T} M^{(t)} = K-A$), yielding the final integrated candidate pool $\mathcal{E}_{\text{ret}} = \mathcal{E}_p^{(T)}$.

As the refinement hops proceed, the trajectory of $q^{(t)}$ must continuously escape the semantic gravity of $\mathcal{E}_p^{(t-1)}$ to discover novel information gaps, thereby maximizing the density of valid evidence within a limited candidate budget $K$.

\section{Methodology}
While our framework executes implicit trajectory steering within a shared latent space, managing this embedding topology reveals foundational structural anomalies. 

First, multi-hop search paths suffer from \textbf{Entity-Centric Bias}, where dominant entities within the initial query $q$ overwhelm the embedding space, fetching redundant assets with diminishing returns (see Figure~\ref{fig:intro}). While breaking this trap necessitates a transition from additive to a subtractive query reformulation, executing such latent feature purification introduces a secondary bottleneck: the \textbf{Invisible Modality Gap}. Because semantic redundancy often transcends modalities, where information in one format (e.g., a film script) may offer the same utility as another (e.g., a corresponding video frame), a disjointed embedding space prevents the retriever from detecting covered information. Consequently, precise cross-modal alignment is a strict prerequisite for effective feature subtraction.

To resolve these limitations, GRAIL operates in two sequential stages: 
(1) \textbf{Unified Semantic Alignment (\ref{subsec:alignment})} synchronizes multimodal vector topologies into a consistent embedding space to enable cross-format redundancy evaluation.
(2) \textbf{Gap-aware Search Mechanism (\ref{subsec:biencoder})} balances acquiring novel complementary information with preserving the original query relevance, suppressing redundant context features from $\mathcal{E}_p$ while maintaining the required logical reasoning path for $q$.

\subsection{Unified Semantic Alignment} \label{subsec:alignment}

The first stage establishes a unified semantic space of text, tables, and images. This alignment is a functional prerequisite for the subsequent bi-encoder retrieval, ensuring that evidence from any modality can be compared and operated upon within a consistent vector space.

\paragraph{Alignment Strategies} We evaluate four strategies, categorized into two query-independent and two query-dependent options.

Query-independent alignments focus on the internal cohesion of the evidence set $\mathcal{E}^*$: (a) \textit{Centroid-based} minimizes the distance between the mean vector of $\mathcal{E}^*$ and its members, while (b) \textit{CLS-based} utilizes the hidden state of the \texttt{[CLS]} token from the concatenated sequence $[\texttt{[CLS]}; e_1; \dots; e_G]$ as a semantic anchor. Conversely, query-dependent approaches incorporate query information into the alignment process: (c) \textit{Query-evidence} explicitly maps $q$ to each individual $e_i \in \mathcal{E}^*$, whereas (d) \textit{Query-set} processes the joint sequence $[q; e_1; \dots; e_G]$ to utilize the query token's final hidden state as the alignment pivot.
 
These alignment strategies differ only in their definition of the anchor representation. For a batch of size $N$, let $anc_x$ be the designated anchor embeddings for $x$-th sample. We optimize an InfoNCE loss to maximize the semantic agreement between this anchor-evidence pairs:
\begin{equation}
    \mathcal{L}_{\text{align}} = -\frac{1}{N} \sum_{x=1}^{N} \frac{1}{|\mathcal{E}_x^*|} \sum_{e \in \mathcal{E}_x^*} \log \frac{\mathcal{P}^x(e)}{\mathcal{P}^x(e) + \mathcal{N}^x}
\end{equation}
where $\mathcal{P}^x(e)$ and $\mathcal{N}_x$ denote the intra-chain positive alignment score and inter-chain negative contrastive score, respectively:
\begin{equation}
    \mathcal{P}^x(e) = \exp\bigl(\text{sim}(anc_x, e) / \tau\bigr)
\end{equation}
\begin{equation}
    \mathcal{N}^x = \sum_{y \neq x}^{N} \sum_{e' \in \mathcal{E}_y^*} \exp\bigl(\text{sim}(anc_x, e') / \tau\bigr)
\end{equation}
where $\tau$ is the temperature parameter and $\text{sim}(\cdot, \cdot)$ denotes the similarity function (e.g., cosine similarity). 

By switching the formulation of $anc$, this objective scales across all strategies introduced above. For example, query-evidence strategy (c) simply instantiates the anchor as $anc_x = q_x$. This unified formalization provides the multi-modal space with the exact topological precision required before executing the subsequent information subtraction stage.

\subsection{Gap-aware Search Mechanism} \label{subsec:biencoder}
Upon the unified latent space, we build a search mechanism to identify the information gap. Overall process is summarized in Algorithm \ref{alg:gap_aware_search}.

\paragraph{Information Aggregation} To represent the knowledge already acquired from $\mathcal{E}_p$, compute a existing knowledge vector $h_{\text{ctx}}$:
\begin{equation}
    h_{\text{ctx}} = \text{LN}\left(\sum_{e_i \in \mathcal{E}_p} \alpha_i e_i\right)
\end{equation}
where the attention weight $\alpha_i$ is defined as:
\begin{equation}
    \alpha_i = \frac{\exp((e_i \cdot q)/ \sqrt{d})}{\sum_{e_j \in \mathcal{E}_p} \exp((e_j \cdot q) / \sqrt{d})}
\end{equation}
where $d$ is the dimension of the shared latent space. 

\begin{algorithm}[t]
\caption{Gap-aware Subtractive Search}
\label{alg:gap_aware_search}
\begin{algorithmic}[1]
\REQUIRE Query $q$, Partial evidence set $\mathcal{E}_p$
\ENSURE Refined query vector $q_{\text{gap}}$, Dynamic temperature $\tau_{\text{dyn}}$
\STATE $V_p \gets \{E(e_i) \mid e_i \in \mathcal{E}_p\}$
\STATE $h_{\text{ctx}} \gets \text{Attention}(q, V_p)$
\STATE $g \gets \sigma(W_{\text{gap}}[q; h_{\text{ctx}}])$
\STATE $q_{\text{gap}} \gets q - g \cdot \mathcal{F}(h_{\text{ctx}})$
\STATE $\tau_{\text{dyn}} \gets \exp(\log \tau_{\text{base}} + \text{MLP}([q; h_{\text{ctx}}]))$
\RETURN $q_{\text{gap}}, \tau_{\text{dyn}}$
\end{algorithmic}
\end{algorithm}

\paragraph{Context-Subspace Removal} To isolate the missing information, we generate the updated query intent via an implicit context-subtractive rewriting of $q$ relative to $h_{\text{ctx}}$. Unlike the unconstrained trajectory expansion of conventional additive search ($q + h_{\text{ctx}}$), our learnable gap gate $g = \sigma(W_{\text{gap}}[q; h_{\text{ctx}}])$ dynamically modulates the steering intensity based on the accumulated context:
\begin{equation}
    q_{\text{gap}} = q - g \cdot \left( \frac{q \cdot h_{\text{ctx}}}{\|h_{\text{ctx}}\|^2} \right) h_{\text{ctx}}
\end{equation}
where $W_{\text{gap}} \in \mathbb{R}^{1 \times 2d}$ and $q_{\text{gap}}$ represents the purified query intent for missing evidence. By design, $g$ uniformly scales the overlapping context to mitigate semantic redundancy without distorting the embedding topology.



\paragraph{End-to-End Joint Optimization} 
All components are optimized jointly and end-to-end without any auxiliary losses or separate heuristics. Because every operation in the gap-aware search mechanism is fully differentiable, the retrieval success signal directly backpropagates to steer these parameters. Specifically, given a training batch of size $N$, let $q_{\text{gap}}^x$ be the dynamically steered request vector for the $x$-th query sample, and $e_t^x$ be its corresponding desired evidence. The model is optimized using the multiple negatives ranking loss formulated as:
\begin{equation}
    \mathcal{L}_{\text{InfoNCE}} = -\frac{1}{N} \sum_{x=1}^{N} \log \frac{\mathcal{P}_{\text{nce}}^x(e_t^x)}{\mathcal{P}_{\text{nce}}^x(e_t^x) + \mathcal{N}_{\text{nce}}^x}
\end{equation}
where $\mathcal{P}_{\text{nce}}^x(e_t^x)$ and $\mathcal{N}_{\text{nce}}^x$ denote the positive retrieval score and the negative contrastive score for the $x$-th sample, respectively:
\begin{equation}
    \mathcal{P}_{\text{nce}}^x(e) = \exp\bigl(\text{sim}(q_{\text{gap}}^x, e) / \tau_{\text{dyn}}^x\bigr)
\end{equation}
\begin{equation}
    \mathcal{N}_{\text{nce}}^x = \underbrace{\sum_{y \neq x}^{N} \mathcal{P}_{\text{nce}}^x(e_t^y)}_{\text{In-batch negatives}} + \underbrace{\sum_{z \in \mathcal{Z}^x} \mathcal{P}_{\text{nce}}^x(z)}_{\text{Hard distractors}}
\end{equation}
where $\mathcal{Z}^x$ represents the set of hard multimodal distractors. The retrieval score is calibrated using a dynamic temperature $\tau_{\text{dyn}}^x$, which automatically regulates how aggressively the network learns from each context.

\subsection{Task-Adaptive Hybrid Framework} 
The optimal strategy for retrieval fundamentally diverges across different question categories. 
For example, certain queries benefit from additive reinforcement ($q_{\text{add}}$) to consolidate ongoing context, while others strictly demand subtractive query steering ($q_{\text{gap}}$) to isolate missing information. 
To exploit these distinct operational advantages, we introduce a \textit{Task-Adaptive Hybrid Framework} governed by a front-end query selector.
By analyzing the incoming query $q$, a lightweight selector $\Phi(q) \in \{0, 1\}$ pre-determines the most effective specialist framework before executing the document search:
\begin{equation}
    q_{\text{ret}} = \begin{cases}
        q_{\text{add}}, & \text{if } \Phi(q) = 0 \;\; \\
        q_{\text{gap}}, & \text{if } \Phi(q) = 1 \;\;
    \end{cases}
\end{equation}
 
The selector $\Phi(q)$ is a Transformer-based sequence encoder topped with a linear classification layer. Independent of the retrieval training, the selector is trained in a self-supervised manner on a routing dataset $\mathcal{D}_{\text{route}} = \{(q, y)\}$. To construct pseudo-ground-truth labels without relying on heuristic task metadata, we evaluate both specialist models ($q_{\text{add}}$ and $q_{\text{gap}}$) on the validation split and assign y, based on their empirical retrieval performance:

\begin{equation}
    y = \begin{cases}
        0, & \text{if } \text{Rank}_{\text{add}}(e_t) < \text{Rank}_{\text{gap}}(e_t) \;\; \\
        1, & \text{otherwise}
    \end{cases}
\end{equation}

where $\text{Rank}(e_t)$ denotes the retrieval rank of the target gold evidence $e_t$.
The selector is optimized using standard cross-entropy loss to predict which paradigm yields a superior ranking trajectory for a given query. Once trained, the selector parameters are frozen. During inference, this gate acts as a deterministic paradigm switch, structurally enabling task-specialized synergy.

\section{Experiments}
\subsection{Experimental Setup}
\subsubsection{Dataset}
We utilize the \textbf{MultimodalQA} benchmark \cite{talmor2021multimodalqa}; its corpus and dataset statistics are detailed in Appendix~\ref{subsec:app_multimodalQA_stats} (Table~\ref{tab:corpus_stats} and Table~\ref{tab:task_split_stats}). (See Appendix~\ref{sec:app_webqa} for WebQA dataset evaluation.)

To evaluate the model's ability to identify the information gap, we reformulate the original MultimodalQA into the Evidence Set Completion task. Based on the gold evidence sets $\mathcal{E}^*$, we filter for instances where $|\mathcal{E}^*| \geq 2$. Specifically, we expand each question-evidence set pair into multiple test instances by iteratively withholding each evidence item $e_i \in \mathcal{E}^*$ as the \textbf{missing evidence} ($e_t = e_i$) while providing all other items in the set as the \textbf{context} ($\mathcal{E}_p = \mathcal{E}^* \setminus \{e_t\}$). This exhaustive leave-one-out approach ensures that the model is evaluated on its ability to retrieve any missing component $e_t$ from the full corpus $\mathcal{C}$ regardless of the provided context.

\subsubsection{Baselines}
Unless specified otherwise, all models are evaluated under the \textit{Query-evidence} alignment (see Appendix~\ref{subsec:app_implementation_details} for full implementation details). We compare GRAIL against several competitive alternatives, ranging from standard dense retrieval to large language model (LLM) query regeneration:

\begin{itemize}[leftmargin=*]
    \item \textbf{Query Only}: A standard dense bi-encoder (DPR-style \cite{karpukhin-2020-dense}) that retrieves evidence solely relying on the original query $q$.
    \item \textbf{Additive}: The search vector is updated by adding the query and context embeddings, analogous to PRF / Rocchio-style relevance feedback \cite{rocchio-1971-relevance} in dense retrieval.\looseness=-1
    \item \textbf{Tuned Rocchio}: A classical relevance feedback baseline \cite{rocchio-1971-relevance} that subtracts negative noise ($q_{\text{ret}} = \text{Norm}(q + \beta R_{\text{mean}} - \gamma N_{\text{mean}})$). Here, $R_{\text{mean}}$ and $N_{\text{mean}}$ represent previous gold evidence and top-$U$ non-context noise, respectively ($\beta=0.25, \gamma=0.25, U=10$).\looseness=-1
    \item \textbf{LLM-based Query Regeneration}: We adapt text-based multi-hop methods to multimodal settings using gpt-4o-mini \cite{gpt-4o-mini}:
    \begin{itemize}[leftmargin=10pt]
        \item \textbf{Closed-Book Generation}: The LLM regenerates a search query based solely on $q$, evaluating the baseline performance of internal parametric knowledge.\looseness=-1
        \item \textbf{IRCoT}: Adapted from IRCoT \cite{trivedi-etal-2023-interleaving}, the LLM generates a reasoning trace given $q$ and $\mathcal{E}_p$, where visual assets are fed as raw images.\looseness=-1
        \item \textbf{Answer-Augmented Retrieval}: The LLM generates an intermediate answer from the context, which is then concatenated to the query ($\text{ans} \oplus q$). \looseness=-1
    \end{itemize}
\end{itemize}
\subsubsection{Evaluation Metrics} \label{subsubsec:metrics}
To evaluate retrieval quality and dissect the inner mechanics, we use two primary performance metrics and three diagnostic metrics.

\paragraph{Primary Performance Metrics}
For Evidence Set Completion (\ref{subsec:evidence_comp}), we utilize \textbf{Recall@K}, measuring whether the single missing target evidence $e_t$ is retrieved. (R@$K = \mathbb{I}(\text{rank}(e_t) \le K)$.)
For Sequential Pool Construction (\ref{subsec:pool_const}), we employ \textbf{Set-Recall@K} to quantify the coverage of the candidate pool $\mathcal{E}_{\text{ret}}$ against the gold evidence set $\mathcal{E}^*$:
\begin{equation}
    \text{Set-Rec}@K (\mathcal{E}_{\text{ret}}) = \frac{|\mathcal{E}_{\text{ret}} \cap \mathcal{E}^*|}{|\mathcal{E}^*|}.
\end{equation}

\paragraph{Diagnostic Metrics}
First, to quantify the semantic trajectory of retrieval vectors and empirically diagnose the phenomenon of semantic anchoring, we introduce the \textbf{Semantic Escape Delta ($\Delta_{\text{esc}}$)}:
\begin{equation}
    \Delta_{\text{esc}} = \text{sim}(q', e_t) - \max_{e_i \in \mathcal{E}_p} \text{sim}(q', e_i).
\end{equation}
A positive $\Delta_{\text{esc}}$ indicates that the steered query vector has successfully broken out of the semantic gravity of the satisfied context $\mathcal{E}_p$ toward the unexplored logical void of the missing target $e_t$.

Second, to evaluate the framework's capability to maximize the information density of the limited candidate pool, we introduce the \textbf{Rank Jump ($Jump$)}. Formally, let $\mathcal{E}_{\text{qo}}$ denote the top-$K$ evidences retrieved solely by the raw query $q$ (corresponding to the $\mathcal{E}_{\text{base}}$ size of K). The subset of newly included target assets $\mathcal{S}$ is formulated as:
\begin{equation}
    \mathcal{S} = (\mathcal{E}_{\text{ret}} \cap \mathcal{E}^*) \setminus \mathcal{E}_{\text{qo}}
\end{equation}
The Rank Jump for a given query $q$ is computed as the mean baseline rank of these elevated assets:
\begin{equation}
    Jump(q) = \frac{1}{|\mathcal{S}|} \sum_{e_i \in \mathcal{S}} \text{rank}_{\text{qo}}(e_i)
\end{equation}
where $\text{rank}_{\text{qo}}(e_i)$ represents the initial rank of asset $e_i$ under the Query Only retrieval. A higher $Jump$ score empirically demonstrates that the model effectively surfaces informative, low-ranked assets into the top-$K$ slots, thereby constructing a highly efficient, information-dense candidate pool.

Third, to quantify the self-purifying capacity under adversarial initial states, we introduce the \textbf{Noise-Resilience Margin ($NRM$)}. We isolate a subset of highly challenging queries, $\mathcal{Q}_{\text{n}}$, where $\mathcal{E}_{\text{base}}$ contains absolutely zero gold items ($\mathcal{E}_{\text{base}} \cap \mathcal{E}^* = \emptyset $). For this 100\% noisy subset, $NRM$ measures the net recall gain achieved by the retrieval model over the Query Only alternative:
\begin{equation}
\resizebox{\columnwidth}{!}{%
$NRM = \frac{1}{|\mathcal{Q}_{\text{n}}|} \sum_{q \in \mathcal{Q}_{\text{n}}} \Big( \text{Set-Rec}@K(\mathcal{E}_{\text{ret}}) - \text{Set-Rec }@K(\mathcal{E}_{\text{qo}}) \Big)$%
}
\end{equation}
A positive $NRM$ confirms model's self-correcting capability; the model successfully limits error propagation and prevents the query from veering off in meaning, navigating back to the target assets even when starting from a completely noisy context.

\subsection{Specialization: Anchoring vs. Purification}

\begin{figure}[t]
  \includegraphics[width=\columnwidth]{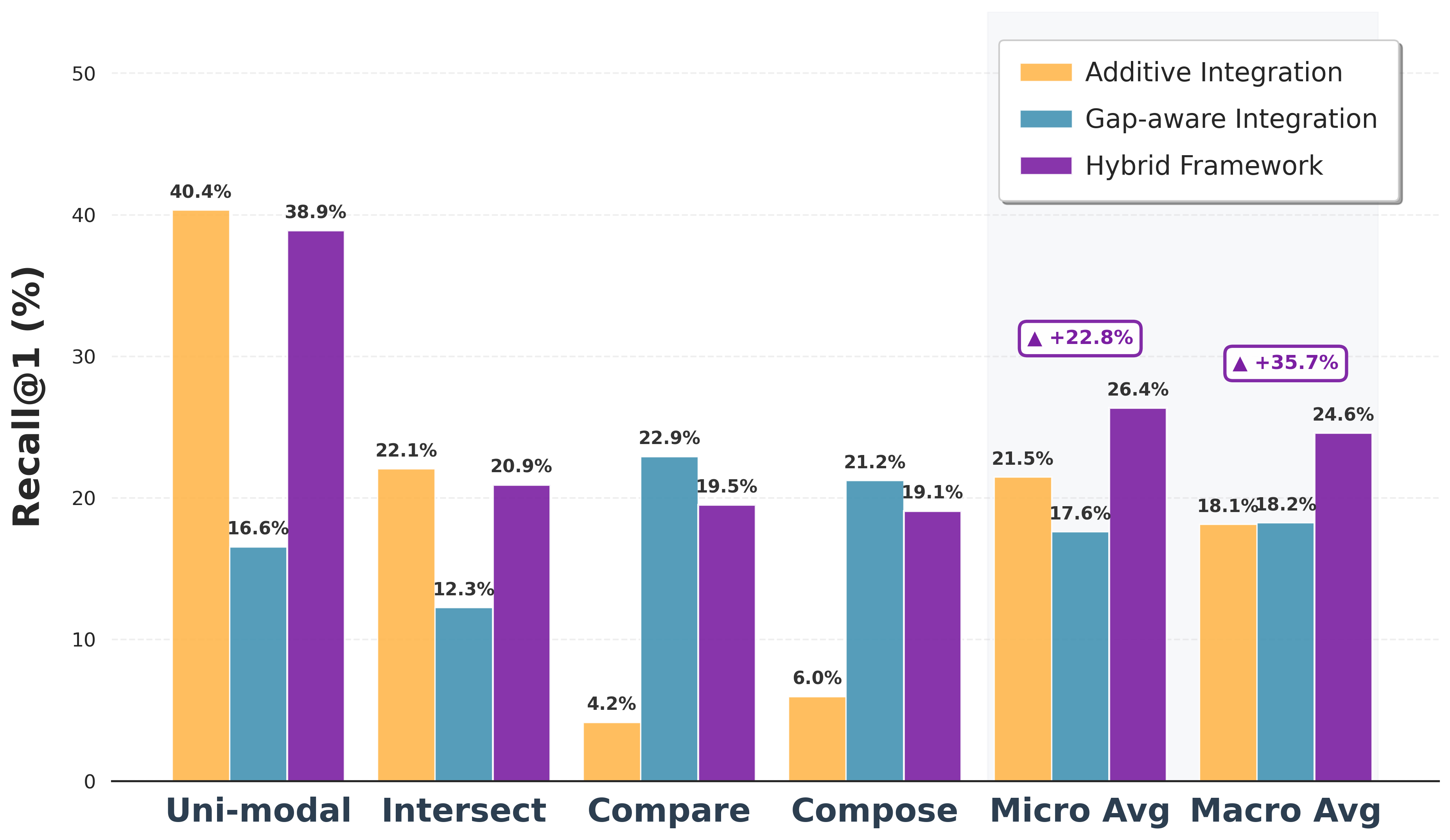}
  \caption{\textbf{Hybrid framework leveraging complementary specialists in MultimodalQA.} By dynamically routing each query to the optimal specialist, our hybrid framework achieves a "best-of-both-worlds" performance (+35.7\% macro avg. gain over Additive) in MultimodalQA test split.}
  \label{fig:hybrid}
\end{figure}

\paragraph{Divergent Performance Patterns} Table~\ref{tab:main_results} and Figure~\ref{fig:hybrid} reveal a clear functional decoupling: Additive integration and Rocchio feedback excel in \textit{Uni-modal} and \textit{Intersect} tasks by acting as a ``topic reinforcer'' within localized semantic domains, but fails significantly in \textit{Compare} and \textit{Compose} tasks requiring entity transitions or cross-modal leaps (see Appendix~\ref{subsec:app_qtype_alignment} for further analysis). In these reasoning-heavy settings, the conventional baselines suffers from \textbf{semantic anchoring}, where the context locks the search vector and blocks the transition to distinct targets. Conversely, GRAIL dominates these tasks, outperforming the baseline by nearly threefold (20.08\% vs. 7.44\% in \textit{Compose}).

\begin{table}
\centering
\footnotesize 
\setlength{\tabcolsep}{1.5pt} 
\caption{\textbf{Evidence Set Completion performance across question types in MultimodalQA.} While the additive and Rocchio baselines performs well on localized aggregated tasks (\textit{Uni-modal}, \textit{Intersect}), GRAIL achieves clear gains on compositional reasoning tasks (\textit{Compose}, \textit{Compare}), 
additionally supported by the positive $\Delta_{\text{esc}}$.
}
\label{tab:main_results}
\begin{tabular}{ll cccc c}
\toprule
\textbf{Type} & \textbf{Method} & \textbf{R@1} & \textbf{R@5} & \textbf{R@10} & \textbf{R@20} & \textbf{$\Delta_{\text{esc}}$} \\ 
\midrule
\multirow{3}{*}{Uni-} & \textbf{Additive} & 37.67 & \textbf{61.00} & \textbf{70.38} & \textbf{77.84} & $-$0.139 \\
 & Rocchio & \textbf{40.91} & 59.09 & 69.21 & 75.60 & $-0.109$ \\  
 & GRAIL & 20.91 & 33.44 & 40.83 & 49.05 & $-0.013$ \\  
\midrule
\multirow{3}{*}{Intersect} & Additive & \textbf{23.00} & 45.71 & 52.81 & 61.90 & $-$0.111 \\
 & \textbf{Rocchio} & 22.90 & \textbf{46.40} & \textbf{54.39} & \textbf{63.47} & $-$0.104 \\
 & GRAIL & 14.12 & 31.69 & 39.29 & 47.98 & $-$0.066 \\ 
\midrule
\multirow{3}{*}{Compare} & Additive & 4.28 & 12.59 & 16.63 & 24.47 & $-$0.245 \\
& Rocchio & 9.26 & 18.29 & 23.75 & 29.69 & $-$0.177 \\
 & \textbf{GRAIL} & \textbf{17.81} & \textbf{32.54} & \textbf{39.43} & \textbf{46.79} & $+$0.041 \\ 
\midrule
\multirow{3}{*}{Compose} & Additive & 7.44 & 19.38 & 26.90 & 33.10 & $-$0.292 \\
& Rocchio & 10.16 & 24.81 & 32.48 & 41.09 & $-$0.190 \\
 & \textbf{GRAIL} & \textbf{20.08} & \textbf{35.81} & \textbf{43.41} & \textbf{49.77} & $+$0.076 \\ 
\bottomrule
\end{tabular}
\end{table}

\paragraph{Overcoming Semantic Gravity} As shown in Table \ref{tab:main_results}, the Additive model's $\Delta_{\text{esc}}$ consistently drops into negative values, indicating that the query representation is swallowed by the semantic gravity of the satisfied context. Mathematically, simple addition cannot generate a search signal orthogonal to what is already retrieved. Conversely, GRAIL executes successful \textbf{semantic purification}, maintaining a positive escape margin ($+0.076$ in \textit{Compose}). By subtracting redundant projections, it creates a repelling force that drives the retrieval vector into the remaining information gap.

\paragraph{The Hybrid Framework}\label{par:hybrid} This task-specific specialization highlights the necessity of a hybrid framework that routes queries based on reasoning requirements. We train two distinct specialist--each optimized exclusively for their respective top-performing question type--and integrate them via a learned query router, which achieves an accuracy of 84.65\% (F1-score: 0.81). Consequently, this dynamic assignment yields a \textbf{+4.9pp gain} in micro average R@1 (Figure~\ref{fig:hybrid}).

\subsection{Effectiveness of Implicit Query Steering}
Table~\ref{tab:query_regeneration} compares our implicit update models with explicit baselines relying on LLM-generated text queries. While methods like IRCoT attempt to trace human-like reasoning, they suffer from a linguistic bottleneck. In multimodal settings, translating high-dimensional visual features into discrete text tokens triggers semantic loss; rich, non-verbal cues are compressed into a few keywords, failing to capture the subtle relational information. Furthermore, the poor performance of Closed-Book and IRCoT--even fall behind the Query Only baseline--demonstrates that explicit text generation can act as a \textbf{semantic distractor}. Inaccurate textual synthesis misdirects the search agent, proving that implicit vector steering directly within the embedding space is fundamentally more robust than error-prone textual translations -- with our Hybrid framework ultimately achieving top retrieval performance (22.21\% R@1 and 53.80\% R@20). For qualitative case studies, see Figure~\ref{fig:intro} and Appendix~\ref{sec:app_case_study}; for inference cost details, see Appendix~\ref{sec:app_cost}.

\begin{table}[t]
  \centering
  \footnotesize
  \setlength{\tabcolsep}{3.5pt}
  \caption{\textbf{Implicit vs. Explicit query modeling performance and latency.} We evaluated macro-averaged recall across the MultimodalQA validation split (2,724 data, $\mathtt{q\_type} \in $\{Compare, Compose, Intersect\}). Our GRAIL \& Hybrid consistently outperform both the additive baseline and explicit query regeneration baselines.}
  \label{tab:query_regeneration}
  \begin{tabular}{l cccc @{\hspace{5pt}} c}
    \toprule
    \textbf{Method} & \textbf{R@1} & \textbf{R@5} & \textbf{R@10} & \textbf{R@20} & \textbf{Latency} \\
    \midrule
    Query Only & 16.24 & 32.68 & {40.51} & 48.54 & 2--5 ms \\
    \midrule
    \multicolumn{6}{l}{\textit{\textbf{Explicit Query Generation}}} \\ 
    \midrule
    Closed-Book & 13.78 & 26.89 & 34.65 & 42.69 & 0.6--1.0 s \\
    IRCoT & 13.42 & 24.87 & 30.18 & 36.21 & 1.1--1.6 s \\
    Ans-Augmented & 16.54 & 33.24 & 39.96 & \underline{48.86} & 1.2--1.8 s \\
    \midrule
    \multicolumn{6}{l}{\textit{\textbf{Implicit Vector Steering}}} \\ 
    \midrule
    Additive & 11.57 & 25.89 & 32.11 & 39.82 & 2--5 ms \\
    GRAIL & \underline{17.55} & \underline{33.61} & \underline{40.81} & 47.50 & 3--6 ms \\
    \midrule
    \textbf{Hybrid (ours)} & \textbf{22.21} & \textbf{39.94} & \textbf{46.90} & \textbf{53.80} & 3--6 ms \\
    \bottomrule
  \end{tabular}
\end{table}


\subsection{Achieving Extended Effective Top-$K$}
\label{subsec:effective_topK}
To evaluate practical efficiency in a realistic pipeline, we analyze sequential pool construction strategies where the evidence pool is built from scratch without relying on oracle context. 
Table~\ref{tab:iteration_strategies} breaks down these behaviors for $K=10$. Each strategies denotes the step-wise retrieval slice sizes (e.g., \texttt{3+7} indicates retrieving 3 assets as the base, then 7 in the next turn).
\begin{table}[t]
\centering
\footnotesize
\setlength{\tabcolsep}{2pt}
\caption{\textbf{Granular efficiency under various step configurations ($K=10$).} Performance changes are relative to the Query Only baseline; Jump values are reported as mean (90th percentile). Results demonstrate that sequential GRAIL outperforms both static additive interpolation and single-shot retrieval, effectively compensating for the scarcity of initial context.}
\label{tab:iteration_strategies}
\begin{tabular}{lccc}
\hline
\textbf{Steps} & \textbf{Set-Rec@10} & \textbf{Jump (90\%)} & \textbf{NRM}\\ \hline
\textit{\textbf{Query Only}} &                  &                     & \\ \hline
\texttt{10} & 44.69\% & -- & $\mathcal{E}_{\text{base}} \cap \mathcal{E}^* = \emptyset $ \\ \hline
\multicolumn{4}{l}{\textit{\textbf{Additive}}} \\ \hline
\texttt{5+5} & $-$3.33pp & 24.9 (39.6) & $-$4.48pp \\ 
\texttt{3+7} & $-$5.08pp & 29.5 (51.1) & $-$6.24pp \\ \hline
\multicolumn{4}{l}{\textit{\textbf{GRAIL (Ours)}}} \\ \hline
\texttt{8+1+1} & {$+$1.71pp} & 28.4 (51.0) & $+$2.50pp \\
\texttt{8+2} & $+$1.55pp & 30.8 (51.0) & $+$2.22pp \\
\texttt{6+2+2} & \textbf{$+$1.82pp} & 31.9 (53.1) & $+$2.79pp \\
\texttt{5+5} & $+$1.39pp & \underline{32.8 (53.3)} & \textbf{$+$3.46pp} \\
\texttt{4+2+2+2} & $+$1.78pp & 32.2 (54.2) & \underline{$+$3.26pp} \\
\texttt{3+7} & $+$0.27pp & \textbf{33.2 (53.7)} & $+$2.03pp \\
\texttt{3+2+3+2} & {$+$1.76pp} & 30.9 (53.6) & {$+$2.12pp}  \\
\texttt{2*5} & $+$1.71pp & 32.3 (54.2) & {$+$2.76pp} \\
\texttt{1*10} & $+$1.64pp & 30.9 (53.6) & $+$2.12pp  \\ \hline
\end{tabular}
\end{table}
\paragraph{Efficiency Amplification}
The behavior of the Jump metric in Table~\ref{tab:iteration_strategies} demonstrates a substantial expansion of the effective search horizon. Across all configurations, GRAIL successfully retrieves complementary evidence originally ranked between the 30th and 35th positions in the baseline pool (mean Jump $\approx$ 30--34). This dynamic implies that sequential information subtraction allows a compressed pool of size $K=10$ to achieve the coverage and search depth equivalent to a three-fold larger Query Only pool. Consequently, our framework offers a resource-efficient alternative that consistently improves gold evidence coverage within a restricted budget, reducing downstream context overhead. \looseness=-1
\begin{figure}[t]
  \includegraphics[width=\columnwidth]{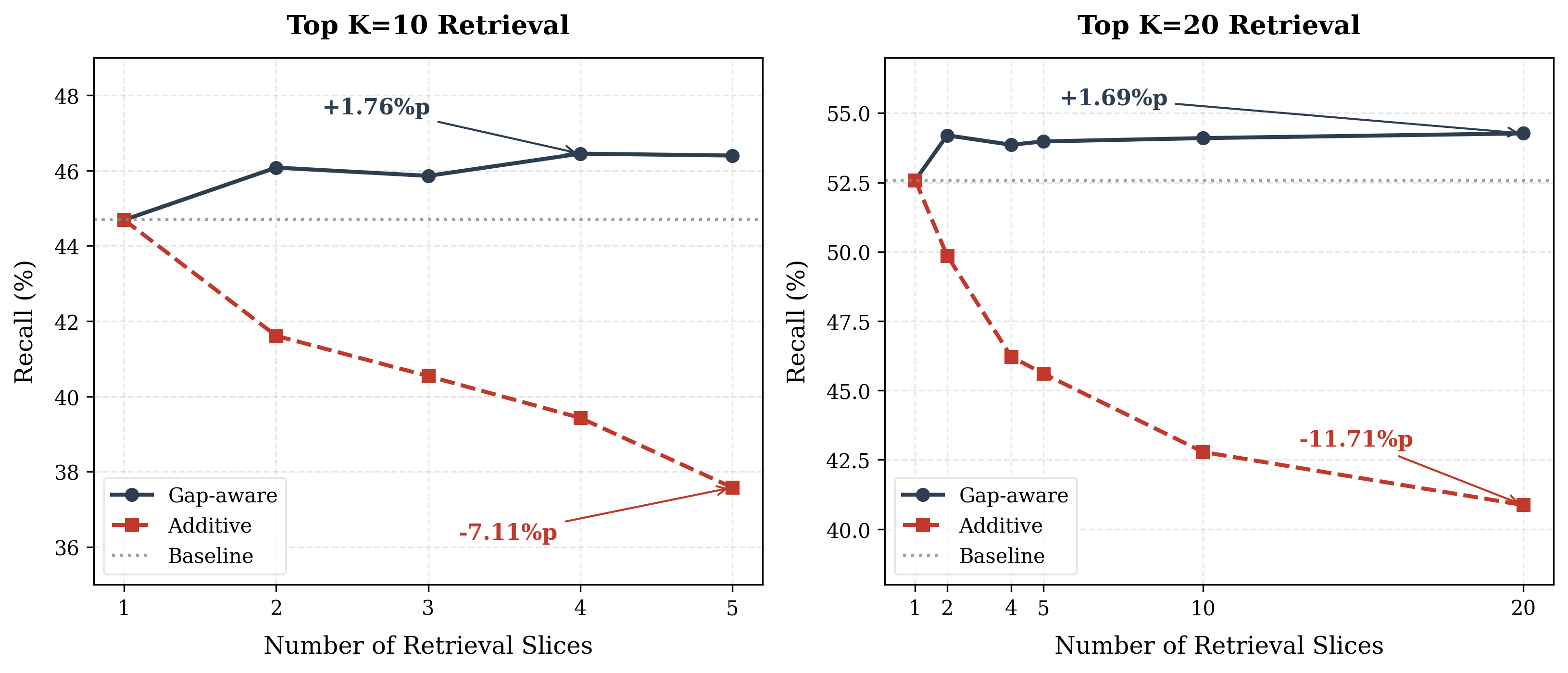}
    \caption{\textbf{Robustness of GRAIL vs. Additive models under iterative slicing.} Set-Recall(\%) is plotted against the number of retrieval slices in MultimodalQA for $K=10$ (left) and $K=20$ (right).
    Although the Additive baseline exhibits strength in specific single-stage question types (Table~\ref{tab:main_results}), it suffers from rapid performance decay under iterative configurations due to error accumulation (see Appendix~\ref{sec:app_iterative} for extended analysis). \looseness=-1}
  \label{fig:pool_comparison}
\end{figure}

\paragraph{Robustness in Noise-Heavy Environments}
The NRM in Table~\ref{tab:iteration_strategies} evaluates framework stability when the initial pool contains zero gold evidence and is entirely polluted by distractors.
Under such noise, traditional paradigms--such as additive rewriting or explicit query regeneration--suffer from error propagation; irrelevant initial documents aggressively mislead subsequent retrieval turns.
In contrast, our framework utilizes projection-based residual subtraction, bypassing fragile textual rewriting.
By tracking uncovered information directly within the embedding space, the model maintains structural robustness even when the initial step is completely blind, achieving a peak robustness gain of \textbf{$+$3.46pp} under \texttt{5+5} strategy.

\subsection{Analysis of Pre-alignment Spaces}
\label{subsec:alignment_performance}

We analyze how cross-modal alignment topologies and training regimes affect latent steering. While query-independent spaces require full fine-tuning (FT) due to coarse feature subtraction, query-dependent approaches perform optimally when frozen, as full FT disrupts pre-aligned semantic boundaries. Crucially, the \textit{Query-evidence} alignment strikes the best balance--capturing precise local contexts while preserving pre-aligned semantic boundaries and yielding a positive $\Delta_{\text{esc}}$. In contrast, \textit{Query-set} alignment suffers from grain-dilution, collapsing localized precision. Consequently, we establish \textit{Query-evidence} as our default topology. We provide full empirical results and deeper theoretical implications in Appendix~\ref{sec:app_alignment_analysis}.

\section{Conclusion}
In this paper, we addressed Semantic Anchoring in MMQA retrieval, where conventional additive methods suffer from entity-centric redundancy. To overcome this, we introduced GRAIL, an implicit context-subtractive query rewriting mechanism operating directly in the shared latent space without error-prone textification. Moreover, our Hybrid Framework dynamically routes queries to task-specific specialists to maximize baseline synergies. Extensive evaluations on MultimodalQA confirm that GRAIL expands the effective search horizon, demonstrates robust noise-resilience, and outperforms explicit query generation baselines.

\section*{Limitations}

While our proposed framework demonstrates significant efficacy and robustness in implicit latent steering for multimodal multi-hop retrieval, we acknowledge several limitations that open avenues for future work.

\paragraph{Fixed Step Scheduling}
In our iterative retrieval experiments (Section~\ref{subsec:effective_topK}), we explored a variety of fixed partitioning strategies (e.g., \texttt{3+2+3+2}, \texttt{5+5}) rather than dynamically adjusting the step sizes during runtime. While this structured scheduling ensures predictable latency and bypasses the computational overhead of an online decision loop, it does not adaptively recalibrate based on the complexity of an individual query. Developing an end-to-end dynamic step controller that infers the optimal stopping criterion per hop remains an open challenge.

\paragraph{Lack of Generative Answer Synthesis}
Our framework focuses strictly on the retrieval and evidence set completion phases of the multimodal multi-hop pipeline. Unlike explicit LLM-based query generation frameworks that end-to-end synthesize final natural language answers, our model is designed as a highly parameter-efficient retrieval accelerator. While it drastically reduces context window overhead and provides clean evidence pools, integrating this implicit steering mechanism with a generative downstream reader to produce comprehensive multimodal answers is left for future engineering.

\paragraph{Scalability to Open-Ended Web Environments}
Although our empirical evaluations comprehensively cover both standard benchmarks and deep multi-hop scenarios (extending to 3+ hops), our projection-based subtraction has primarily been verified within bounded knowledge corpora. Scalability to open-ended, web-scale retrieval environments--where the density of distractors increases exponentially and the modality drift is more severe--remains an active area of investigation.

\section*{Acknowledgments}
This work was supported in part by the National Research Foundation of Korea (NRF) grant (RS-2023-00280883, RS-2023-00222663); by the National Research Foundation, Korea, under project BK21 FOUR(Dept. of Data Science, SNU, No. 5199990914569); by the Korea Institute of Science and Technology Information (KISTI) in 2026 (No. (KISTI)K26L3M1C1), aimed at developing KONI (KISTI Open Neural Intelligence), a large language model specialized in science and technology; and by the Institute of Information \& communications Technology Planning \& Evaluation (IITP) grant funded by the Korea government(MSIT) (RS-2025-02263754, Human-Centric Embodied AI Agents with Autonomous Decision-Making); by grant (25202MFDS003) from Ministry of Food and Drug Safety in 2025; by AI-BIO Research Grant through Seoul National University; Institute of Information \& communications Technology Planning \& Evaluation (IITP) grant funded by the Korea government(MSIT) (No. RS-2025-25442149, LG AI STAR Talent Development Program for Leading Large-Scale Generative AI Models in the Physical AI Domain).

\bibliography{reference}

\appendix
\section{Additional Qualitative Case Study} \label{sec:app_case_study}
\begin{figure*}
  \centering
  \includegraphics[width=\textwidth]{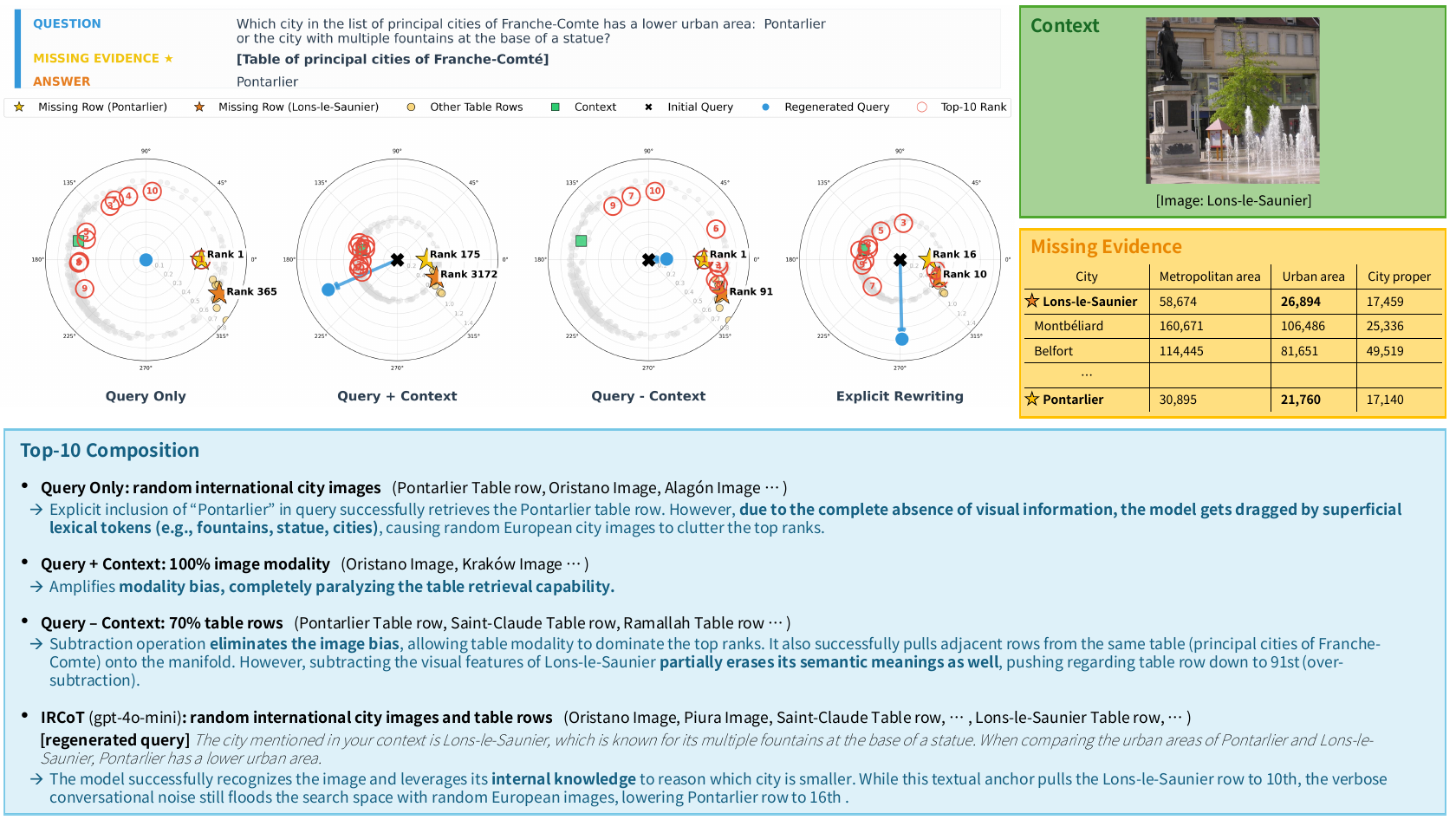}
  \caption{\textbf{Additional Case Study.} Given an image of Lons-le-Saunier, the goal is to fetch the missing table evidence (\textit{the list of principal cities of Franche-Comté}) to complete the evidence set.}
  \label{fig:app_case_study}
\end{figure*}

To further illustrate the practical mechanics of GRAIL, we provide a detailed qualitative visualization of a representative multi-hop retrieval trajectory (Figure~\ref{fig:app_case_study}).

The selected query requires comparing the urban areas of \textit{Pontarlier} and another city depicted in an image featuring a statue and multiple fountains(\textit{Lons-le-Saunier}). To successfully resolve this example, the retriever faces a distinct cross-modal bottleneck: it must identify the visual asset as \textit{Lons-le-Saunier} while simultaneously retrieving the target structured table (\textit{Principal cities of Franche-Comté}) containing the precise metrics for both cities.

When analyzing the latent space and Top-10 retrieval composition, conventional paradigms suffer severely from embedding distortion. For instance, in the \texttt{Query Only} setting, while explicitly searching for ``Pontarlier'' pulls the missing table row of Pontarlier, the complete absence of visual context forces the model to anchor on superficial text keywords such as \textit{fountains, statue, or cities}. This populates the top ranks with irrelevant international city images. Similarly, naively adding the context image embedding (\texttt{Query+Context}) amplifies the modality bias; the search manifold becomes 100\% dominated by image representations, completely paralyzing the model's table retrieval capabilities.

Even with \textbf{Explicit Rewriting} (IRCoT via GPT-4o-mini), a severe confirmation bias emerges. Instead of recognizing that the entire structured frame--the \textit{list of principal cities of Franche-Comté}--is missing and guiding the search toward that global layout, the LLM deduces the final answer using its internal knowledge. It then regenerates an over-confident, verbose query specifically designed to justify this early conclusion. While this predictive shortcut manages to fetch the two local target rows (Lons-le-Saunier at 10th and Pontarlier at 16th), the conversational nature of the generated query introduces unnecessary noise, still retrieving a large number of irrelevant international city images within the Top-10 ranks. More importantly, rather than recognizing the global context of the \textit{Franche-Comté} table as the primary search target, this paradigm remains heavily entity-centric. By locking into a predicted entity instead of capturing the broader table layout, it deprives the model of any opportunity to self-correct when its internal knowledge is flawed, potentially leading to a failure in delivering accurate, evidence-based reasoning.

In contrast, our context-subtractive approach (\texttt{Query-Context}) successfully redirects the query embedding toward the missing evidence. By subtracting the visual features of the satisfied context, GRAIL eliminates the dominating image-modality bias, allowing the table representations to dominate the top ranks (70\% table rows) and pulling adjacent rows from the target Franche-Comté table. However, this case also uncovers a nuanced \textit{over-subtraction} phenomenon. As illustrated in Figure~\ref{fig:app_case_study}, while subtractive steering surfaces the global table structure, erasing the visual features of Lons-le-Saunier partially weakens its entity-level representation, meaning the specific missing row for Lons-le-Saunier is not yet sufficiently pulled into the Top-10 ranks (remaining at the 91st rank). 

Mechanistically, over-subtraction occurs under \textbf{near-collinear geometric alignment} between the context vector ($\mathbf{e}_{\text{ctx}}$) and the target entity vector ($\mathbf{e}_{\text{t}}$) in the latent embedding space. When a satisfied visual context shares high directional similarity or overlapping latent sub-dimensions with the unretrieved target entity ($\mathbf{e}_{\text{ctx}} \cdot \mathbf{e}_{\text{t}} \gg 0$), orthogonal projection or vector subtraction aggressively cancels out shared semantic components. Consequently, subtracting $\mathbf{e}_{\text{ctx}}$ involuntarily penalizes shared relational attributes required to pinpoint $\mathbf{e}_{\text{t}}$.

Rather than a failure, we highlight the potential of an \textit{iterative retrieval} paradigm. Since the subtractive steering successfully anchors the global table layout within the Top-10 embedding space, these newly retrieved structural assets serve as a powerful stepping stone. In subsequent retrieval turns, leveraging this intermediate structural foundation allows GRAIL to naturally recover the weakened local entities, progressively narrowing the semantic gap without introducing conversational or modality bias.

\section{Detailed Experiments Setup}
\subsection{Detailed MultimodalQA Statistics}
\label{subsec:app_multimodalQA_stats}
Detailed statistics of the MultimodalQA benchmark \cite{talmor2021multimodalqa} splits used in this work are summarized in Table~\ref{tab:corpus_stats} and Table~\ref{tab:task_split_stats}.

\begin{table}[h]
\centering
\small
\caption{Statistics of the MultimodalQA corpus.}
\label{tab:corpus_stats}
\begin{tabular}{l|ccc|c}
\hline
\textbf{Modality} & \textbf{Text} & \textbf{Table} & \textbf{Image} & \textbf{Total} \\ \hline
\textbf{Count} & 218,285 & 144,407 & 57,058 & 419,750 \\ \hline
\end{tabular}
\end{table}

\begin{table}[t]
\centering
\small
\setlength{\tabcolsep}{3pt}
\caption{Statistics of the MultimodalQA question type. Unique QIDs are expanded into Total Inst. via our leave-one-out approach.}
\label{tab:task_split_stats}
\begin{tabular}{l|rrr|r}
\hline
\textbf{Q\_Type} & \textbf{Train} & \textbf{Val} & \textbf{Test} & \textbf{Total} \\ \hline
Uni-modal & 3,182 & 365 & 409 & 3,956 \\
Intersect & 1,069 & 161 & 149 & 1,379 \\
Compare & 1,317 & 165 & 159 & 1,641 \\
Compose & 4,964 & 625 & 600 & 6,189 \\ \hline
\textbf{Unique QIDs} & 10,532 & 1,316 & 1,317 & 13,165 \\
\textbf{Total Inst.} & 31,095 & 3,929 & 4,024 & 39,048 \\ \hline
\end{tabular}
\end{table}


\subsection{Implementation Details} \label{subsec:app_implementation_details}
For the unified latent space, we employ BGE-large-en-v1.5 \cite{bge_embedding} to embed questions, text passages, and table rows. Images are embedded using CLIP ViT-B/32 \cite{clip_embedding}. All embeddings are projected into a shared latent space of dimension $d=1024$. We train the retrieval model using the AdamW optimizer with a batch size of 32 for 100 epochs, applying a learning rate of $1 \times 10^{-4}$. The base temperature $\tau_{\text{base}}$ is set to $0.05$. All experiments are conducted on NVIDIA RTX A6000 GPUs. The selector $\Phi(q)$ is instantiated using a \texttt{bert-base-uncased} backbone \cite{devlin-2019-bert} fine-tuned independently for 3 epochs with a learning rate of $1 \times 10^{-5}$.

\section{Detailed Alignment Strategies Analysis}
\label{sec:app_alignment_analysis}

This section extends the empirical evaluation of Section~\ref{subsec:alignment} by providing a comprehensive analysis of the interaction between cross-modal alignment topologies, training regimes (\texttt{Frz} vs. \texttt{FT}), and downstream task complexities.

\begin{figure}[t]
  \includegraphics[width=\columnwidth]{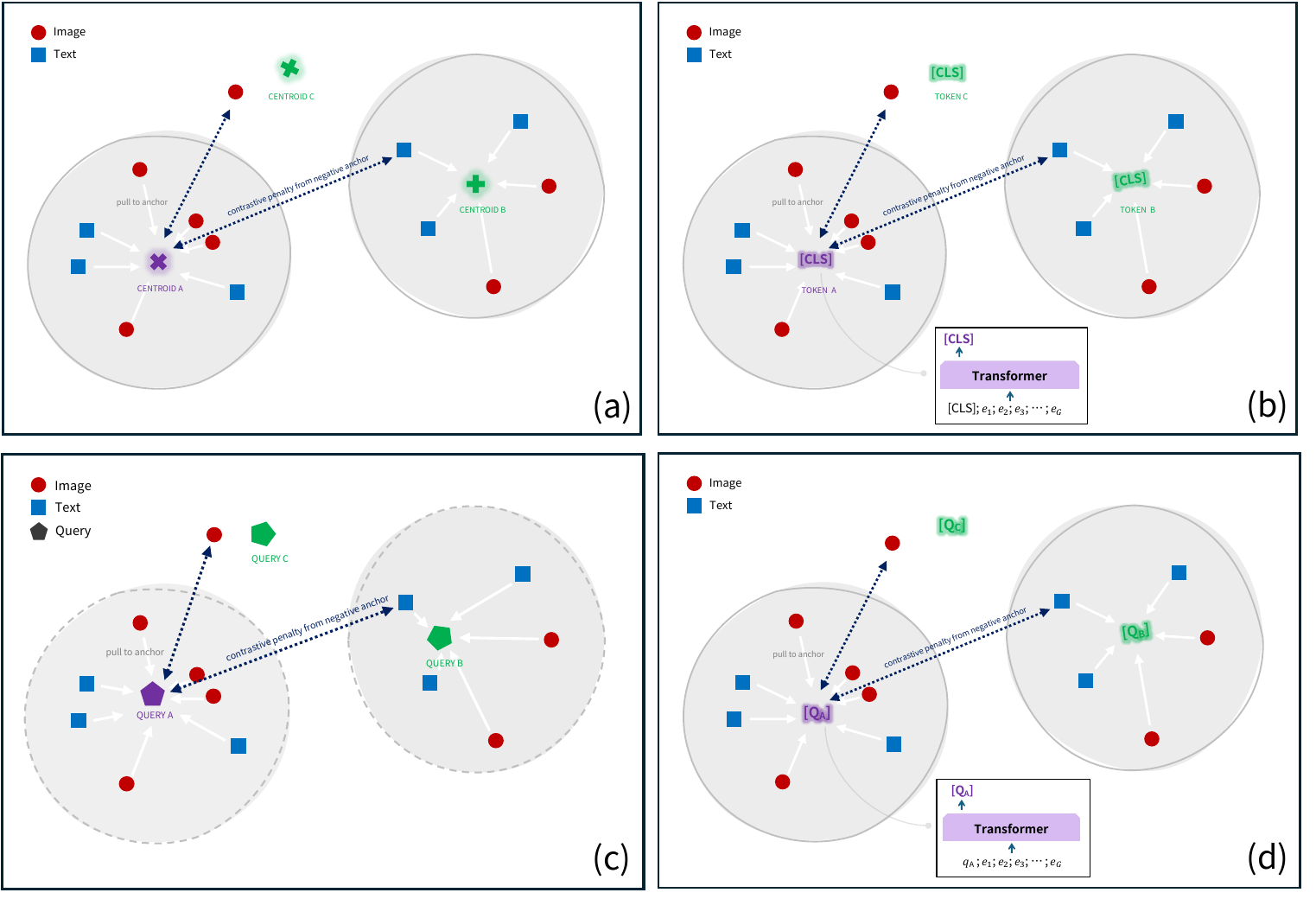}
  \caption{\textbf{Taxonomy of cross-modal semantic alignment.} 
  Top row shows query-independent approaches: 
  (a) Centroid-based alignment,
  (b) CLS-based alignment.
  Bottom row shows query-dependent approaches: 
  (c) Query-evidence alignment,
  (d) Query-set alignment.}
  \label{fig:alignment}
\end{figure}

\subsection{Cross-Examination of Training Regimes}
Table~\ref{tab:all_alignment_set} presents the complete grid search across all alignment strategies and training configurations.

\begin{table}[h]
\centering
\footnotesize
\setlength{\tabcolsep}{1.5pt}
\caption{\textbf{Retrieval performance of gap-aware mechanism across different alignment strategies in MultimodalQA.} Full empirical comparison across all cross-modal topologies under both frozen and fine-tuned training regimes. (\texttt{FT}: full fine-tuning, \texttt{Frz}: frozen backbones).}
\label{tab:all_alignment_set}
\begin{tabular}{llccccc}
\hline
\textbf{Alignment} & \textbf{} & \textbf{R@1} & \textbf{R@5} & \textbf{R@20} & \textbf{R@50} & \textbf{$\Delta_{\text{esc}}$} \\ \hline
None & \texttt{Frz} & 0.81 & 7.05 & 16.47 & 22.35 & {$-$0.094}\\ 
 & \textbf{\texttt{FT}} & \underline{9.04} & \underline{24.97} & 43.24 & 53.93 & \textbf{$+$0.048}\\  \hline
Centroid & \texttt{Frz} & 1.22 & 10.94 & 25.43 & 34.46 & {$-$0.093}\\
& \textbf{\texttt{FT}} & 0.99 & 22.35 & 46.50 & 53.35 & $-$0.162 \\

CLS & \texttt{Frz} & 0.94 & 10.59 & 27.00 & 36.63 & $-$0.088 \\
& \textbf{\texttt{FT}} & 1.22 & 24.03 & \underline{47.03} & 54.67 & $-$0.196 \\ \hline

Query-evidence  & \textbf{\texttt{Frz}} & \textbf{15.63} & \textbf{30.72} & \textbf{47.72} & \textbf{58.36} & \underline{$+$0.008} \\
& \texttt{FT} & 0.03 & 0.13 & 0.20 & 0.36 & $-$0.179 \\
Query-set & {\texttt{Frz}} & 7.64 & {19.60} & {33.90} & {44.74} & $-$0.032\\ 
& \texttt{FT} & 0.92 & 22.22 & 46.55 & \underline{55.18} & $-$0.193 \\ \hline
\end{tabular}
\end{table}

The empirical behavior of query-independent approaches (\textit{None}, \textit{Centroid}, \textit{CLS}) further confirms that optimizing for internal evidence set cohesion alone is insufficient for multi-modal alignment. Relying solely on intra-chain similarity or static aggregation (\textit{Centroid}, \textit{CLS}) lacks a definitive semantic anchor to justify why disparate modalities should co-exist within the same evidence chain. Without a central medium to synthesize these cross-modal fragments, frozen setups (\texttt{Frz}) yield near-zero performance (e.g., R@1 of 0.81\% for \textit{None}, 1.22\% for \textit{Centroid} and 0.94\% for \textit{CLS}), requiring full fine-tuning (\texttt{FT}) to forcefully warp the unconditioned representations into task-relevant boundaries.

In contrast, our query-dependent formulations introduce the query as an explicit semantic medium. Rather than blindly pushing text, tables, and images together, strategies like \textit{Query-evidence} leverage the query to explicitly contextualize why these specific cross-modal segments belong to the same evidence chain.
Consequently, when this already optimized space undergoes full fine-tuning (\texttt{FT}), it triggers severe overfitting and representation drift instead of further adaptation. As evidenced by the \textit{Query-evidence} R@1 dropping drastically from 15.63\% to 0.03\% (with $\Delta_{\text{esc}}$ turning to $-0.179$), full parameter updates break the highly precise, cross-modal arrangements. Therefore, as established in our main experiments (Section~\ref{subsec:alignment_performance}), we evaluate each alignment style under its respective optimal training regime (\texttt{FT} for query-independent and \texttt{Frz} for query-dependent settings) to conduct a follow-up analysis.

\subsection{Alignment Adaptability by Question Type} \label{subsec:app_qtype_alignment}

To analyze how alignment strategies interact with specific question characteristics, Figure~\ref{fig:app_qtype_alignment} breaks down the performance across distinct question categories (Uni-modal, Intersect, Compare, Compose). The results reveal a clear architectural specialization rooted in the type of reasoning required:
\begin{figure}[t]
  \includegraphics[width=\columnwidth]{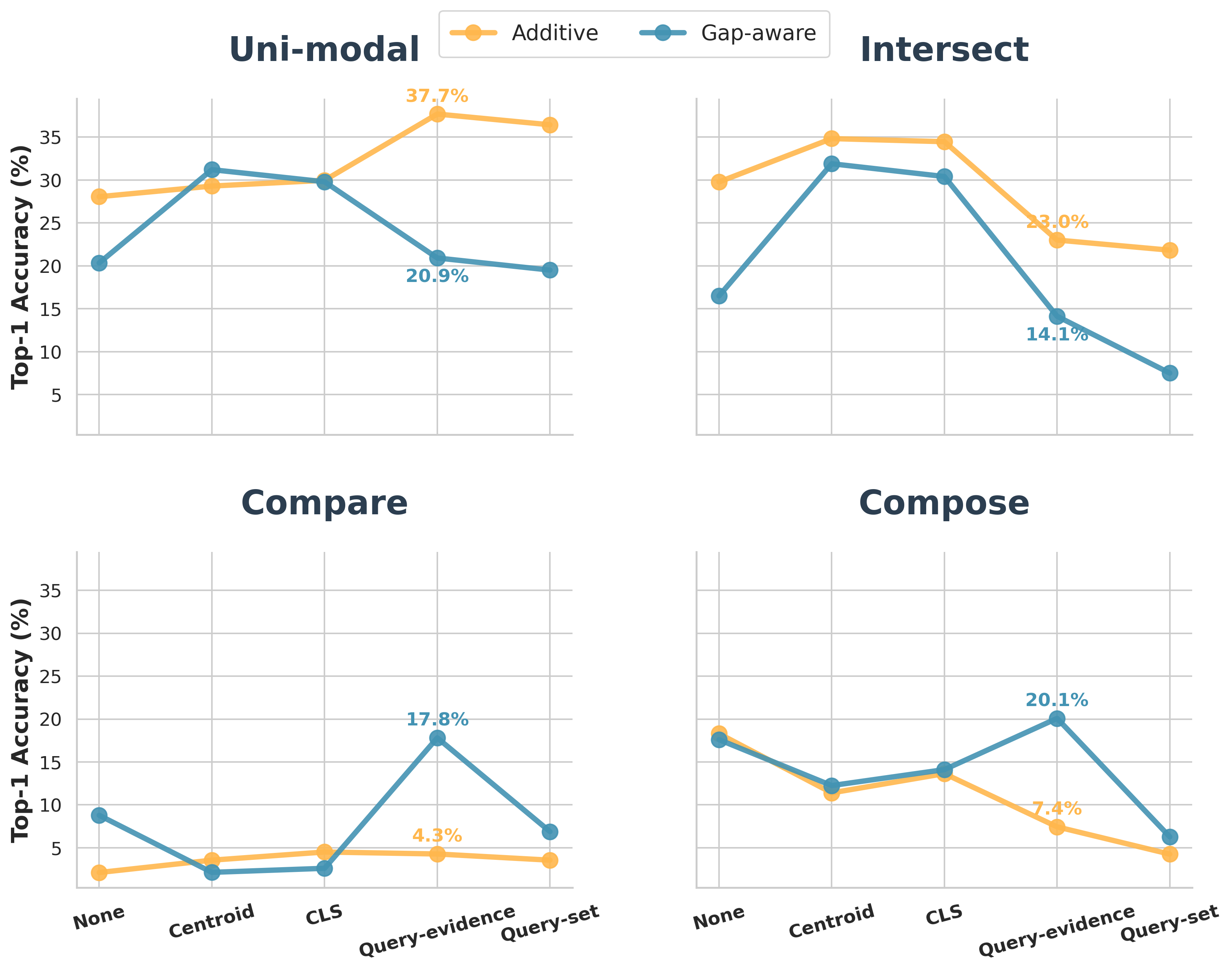}
  \caption{\textbf{Model specialization by task complexity in MultimodalQA.} Additive models excel in Uni-modal lookup, while Gap-aware models dominate reasoning-heavy tasks (Compare, Compose). This performance gap remains consistent across alignment architectures.}
  \label{fig:app_qtype_alignment}
\end{figure}
\begin{itemize}
    \item \textbf{Localized Aggregation Tasks} (Uni-modal, Intersect): The conventional Additive baseline consistently outperforms, peaking at 37.7\% recall in Uni-modal queries. These tasks primarily rely on single-modality lookups or localized information aggregation where representation fusion is sufficient. In these environments, our subtractive steering introduces unnecessary geometric disruption to vectors that are already optimal for concentrated entity matching.
    \item \textbf{Compositional Reasoning Tasks} (Compare, Compose): Our Gap-aware mechanism dominates the baseline, achieving 17.8\% accuracy in \textit{Compare} (vs. 4.3\% for Additive) and 20.1\% in \textit{Compose} (vs. 7.4\% for Additive). Because these tasks demand genuine compositional cross-modal reasoning--where information across disparate modalities and distinct entities must be cross-referenced without redundancy--GRAIL successfully isolates the missing semantic fragments where additive methods fail.
\end{itemize}

By observing the distinct behavior of each alignment strategy across these tasks, we propose several hypotheses regarding how vector space constraints impact retrieval.

Specifically, for \textbf{Uni-modal} queries, the Additive baseline exhibits a sharp upward trajectory that explicitly peaks at the \textit{Query-evidence} stage, reaching its highest accuracy of 37.7\%. Since Uni-modal tasks can be resolved entirely within a single modality, the retrieval space does not require complex multi-modal bridging. In this context, \textit{Query-evidence} provides the clean lexical-to-semantic alignment needed to guide the additive vector update straight toward the target zone within specific modality.

Conversely, for \textbf{Intersect} queries, the Additive architecture achieves its performance plateau at the query-independent \textit{Centroid} and \textit{CLS} stages, but experiences a significant drop down to 23.0\% at the \textit{Query-evidence} stage. We hypothesize that Intersect tasks inherently require capturing the overlapping intersection across disparate modalities. Query-independent pre-alignment naturally compacts these heterogeneous text, table, and image assets into a dense, shared cross-modal cluster beforehand, making it easy for a simple additive update to fuse the multi-modal evidence. However, because the \textit{Query-evidence} strategy focuses strictly on mapping the query to each asset individually, it fails to bind the disparate modalities to one another. This lack of direct cross-modal cohesion isolates the assets from each other, fracturing the dense cluster structure and making localized information aggregation for Intersect queries much more difficult.

Conversely, in the \textbf{Compare} task, while query-independent methods severely degrade retrieval, the query-dependent \textit{Query-evidence} strategy yields a powerful performance spike. This behavior indicates that comparison tasks inherently require query-centric alignment rather than inter-asset correlations. Because the target assets often share no intrinsic relationship other than being requested by the same query, forcing query-free clusters groups them by irrelevant modal or thematic similarities, making it highly counterproductive for reasoning. In contrast, the \textit{Query-evidence} strategy successfully maps each asset directly to the query, providing the clean contrastive space needed to compare distinct entities.

Finally, the \textbf{Compose} task uncovers a unique phenomenon: the unaligned \textit{None} setup yields the highest accuracy for the gap-aware mechanism, while among the pre-aligned alternatives, performance peaks at \textit{Query-evidence}. This behavior likely stems from \textit{representation collapse} during the alignment phase. Because compositional queries require assembling open-ended evidence fragments, forcing a structured alignment objective shifts the representations toward rigid cluster centers. This restriction compresses the semantic flexibility needed to link disparate pieces of evidence. While the unaligned (\textit{None}) space preserves this flexibility intact for subtractive isolation, the item-level granularity of the \textit{Query-evidence} configuration serves as the optimal alternative among the aligned models by minimizing global spatial distortion.

\section{Detailed Analysis on Iterative Retrieval}
\label{sec:app_iterative}

This section provides an extended analysis of our iterative retrieval paradigm, detailing statistical significance via bootstrapping, question-type breakdowns, and downstream QA impacts under sequential pool construction.

\paragraph{Statistical Significance of Step Partitioning}
To ensure the statistical robustness of our iterative framework, Table~\ref{tab:iter_bootstrap} reports the 95\% confidence intervals (CI) and standard errors across different partitioning step sequences on Set-Recall@10. Across all configurations, iterative GRAIL consistently improves upon the static \texttt{Query Only} baseline ($44.69\%$). Notably, the \texttt{6+2+2} partitioning scheme emerges as the most effective setup, achieving the highest Set-Recall@10 of $46.51\%$ ($95\%\text{ CI: } [44.49, 48.57]$) and confirming that stepwise subtractive steering provides reliable gains over noisy historical contexts.

\begin{table}[h]
\centering
\footnotesize
\setlength{\tabcolsep}{1.5pt}
\caption{\textbf{Bootstrap evaluation of sequential pool construction.} Comparison of Set-Rec@10 evaluated via query-level bootstrap resampling (N = 1000). Results report means, 95\% Confidence Intervals(CI), and Standard Errors across different stepwise partitioning sequences in a noisy, realistic sequential retrieval setup.}
\label{tab:iter_bootstrap}
\begin{tabular}{lccc}
\hline
\textbf{Steps} & \textbf{Set-Rec@10} & \textbf{95\% CI} & \textbf{Standard Error} \\ \hline
Query Only & 44.69\% & (42.68, 46.75) & 1.033 \\ 
\texttt{8+2} & 46.24\% & (44.20, 48.29) & 1.041 \\
\texttt{6+2+2} & 46.51\% & (44.49, 48.57) & 1.036 \\
\texttt{3+7} & 44.96\% & (42.96, 47.02) & 1.026 \\
\texttt{3+2+3+2} & 46.45\% & (44.45, 48.49) & 1.026 \\
\hline
\end{tabular}
\end{table}

\paragraph{Scalability across Retrieval Budgets}
\begin{figure}[t]
    \centering
    \includegraphics[width=\columnwidth]{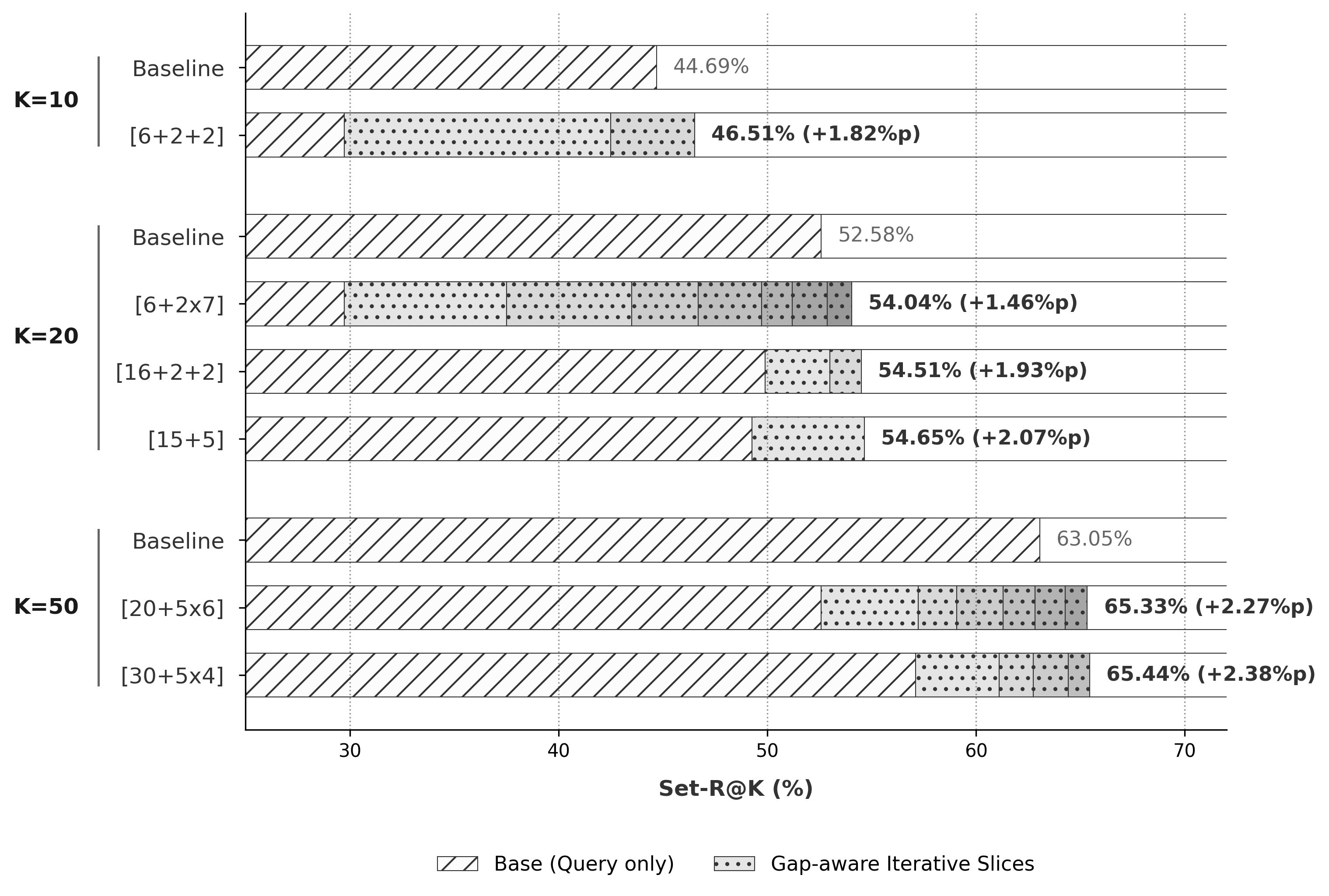}
    \caption{\textbf{Set Recall across varying pool sizes ($K$).} The leftmost dashed segment represents the initial Base (Query Only) retrieval, while subsequent segments illustrate cumulative gains from the sequential execution of iterative GRAIL slices (e.g., \texttt{[3+2]*2} denotes a \texttt{3(base)+2+3+2} step sequence, darkening chronologically). Across all pool sizes, our iterative framework consistently expands retrieval coverage, demonstrating sustained gains even as the retrieval constraints scale.}
    \label{fig:pool}
\end{figure}
As illustrated in Figure~\ref{fig:pool}, our gap-aware iterative strategy yields sustained gains over the baseline across all expanded budget scales ($K \in \{10, 20, 50\}$). By progressively steering the query embedding toward unretrieved semantic space, GRAIL expands retrieval coverage consistently across all pool sizes, proving that subtractive steering does not saturate even under higher recall budgets ($+$1.82pp for Top-10, $+$2.07pp for Top-20, and $+$2.38pp for Top-50).

\paragraph{Noise Isolation across Question Types.}
\begin{figure}[t]
  \includegraphics[width=\columnwidth]{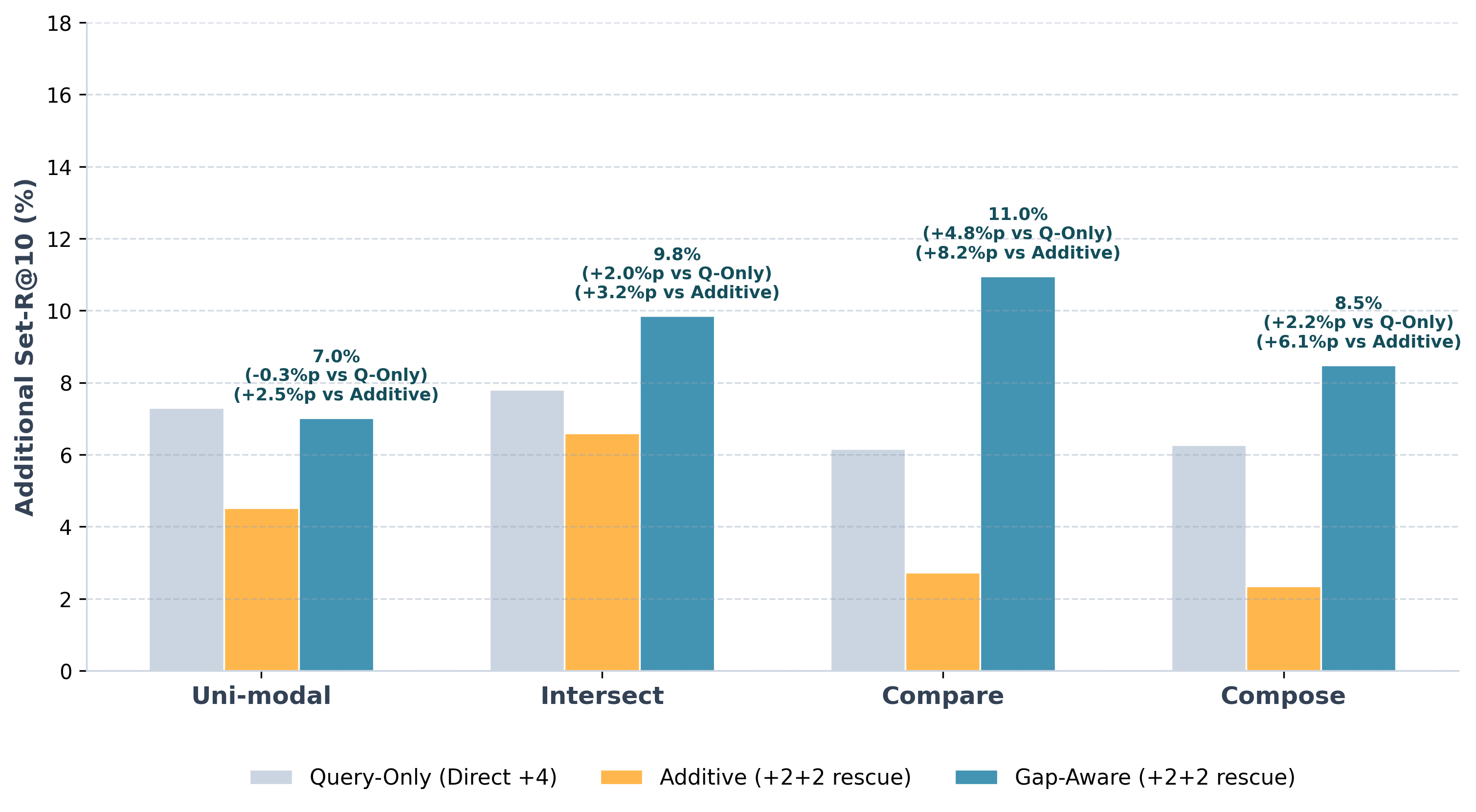}
  \caption{\textbf{Breakdown of additional retrieval gains by question type.} The bars represent the isolated increase in Set-Recall contributed exclusively by the 4 additional evidence pieces retrieved after the initial \textit{Query Only} baseline (Base 6). Note that the baseline recall at Base 6 is omitted, plotting only the net incremental gain captured during the remaining sequential rescue steps (partitioning scheme \texttt{6+2+2}). All values represent the percentage point (ppp) improvement, mapping the distinct behavior of the static \textit{Query Only} baseline, the \textit{Additive} model, and our \textit{Gap-Aware} mechanism across the four question categories of MultimodalQA dataset.}
  \label{fig:q_type_pool}
\end{figure}

To investigate the dynamic behavior of multi-hop retrieval across distinct question categories, Figure~\ref{fig:q_type_pool} illustrates the incremental Set-Recall@10 gains achieved exclusively during the sequential rescue steps (partitioning scheme \texttt{6+2+2}). 
GRAIL delivers the highest net improvement across all multi-modal reasoning categories. Particularly in the \textit{Compare} category, GRAIL outperforms the \textit{Query Only} baseline by 4.8pp and the \textit{Additive} model by \textbf{8.2pp}.

This performance gap stems from the \textit{Additive} model's susceptibility to noise accumulation. In realistic setups starting from unverified \texttt{Query Only} baselines (Sequential Pool Construction), initial retrieval errors are continuously aggregated into the updated vector. Consequently, additive updates distort the search manifold. In contrast, GRAIL's subtractive operator effectively shields the query from accumulated noise, successfully redirecting the search toward the remaining missing evidence.

An identical vulnerability directly accounts for the performance decay observed under iterative retrieval regimes. While the \textit{Additive} approach is known to perform well when initialized with clean, partial gold evidence (Evidence Set Completion), its performance collapses in realistic setups that start from unverified \textit{Query Only} baselines (Sequential Pool Construction). Without the guarantee of gold context, the additive fusion absorbs initial retrieval errors, whereas GRAIL yields stable, superior improvements by strictly isolating the remaining semantic gaps without amplifying noise.

\paragraph{End-to-End Downstream Performance.}
\begin{table}[h]
\centering
\footnotesize
\setlength{\tabcolsep}{1.5pt}
\caption{\textbf{Downstream QA performance.} End-to-End Question Answering metrics (Exact Match and F1) using \texttt{gpt-4o-mini} reasoning module under different iterative retrieval partitioning schemes.}
\label{tab:downstream_score}
\begin{tabular}{lccc}
\hline
\textbf{Steps} & \textbf{Set-Rec@10} & \textbf{QA EM} & \textbf{QA F1} \\ \hline
Query Only & 44.69\% & 33.74 & 30.66 \\ 
\texttt{3+2+3+2} & 46.45\% & 36.02 & 32.71 \\
\texttt{6+2+2} & 46.51\% & 36.40 & 32.96 \\
\hline
\end{tabular}
\end{table}

Finally, Table~\ref{tab:downstream_score} confirms that improved retrieval coverage directly translates into downstream QA performance gains. Rather than acting as a superficial retrieval metric, the elevated Set-Recall directly facilitates downstream reasoning by supplying more complete evidence chains to the LLM. Compared to the \texttt{Query Only} baseline (33.74 EM / 30.66 F1), executing iterative retrieval yields performance gains. Reflecting its top retrieval recall, the \texttt{6+2+2} scheme achieves the highest overall accuracy, reaching \textbf{36.40} EM and \textbf{32.96} F1 score, firmly establishing that GRAIL's gap-aware steering effectively empowers end-to-end multi-modal reasoning.

\section{Ablation Study}
To understand the inner workings of GRAIL, we analyze the behavior of the dynamic gating coefficient ($g$) across different question categories.

\begin{table}[h]
\centering
\footnotesize
\setlength{\tabcolsep}{1.5pt}
\caption{\textbf{Distribution of gating coefficients on MultimodalQA.}
Values are reported as mean (standard deviation) on the MultimodalQA validation set. Higher gate values indicate stronger subtractive vector steering.}
\label{tab:gate_mix_stats}
\begin{tabular}{lc}
\toprule
\textbf{Category} & \textbf{Gate} $g$ \\
\midrule
Uni-model & $0.233\ (0.099)$ \\
Intersect & $0.264\ (0.107)$ \\
Compare   & $0.286\ (0.095)$ \\
Compose   & $0.260\ (0.094)$ \\
\midrule
\textbf{Overall} & ${0.288\ (0.091)}$ \\
\bottomrule
\end{tabular}
\end{table}

\paragraph{Dynamic Gating Activation}
Table~\ref{tab:gate_mix_stats} provides a detailed statistical breakdown of the learned gating parameter $g$. The dynamic gating coefficient maintains an overall average of 0.288, displaying higher activation in complex multi-modal reasoning tasks such as \textit{Compare} (0.286) and \textit{Intersect} (0.264) relative to simpler \textit{Uni-modal} queries (0.233). This confirms that GRAIL actively modulates the intensity of the subtractive operation according to the reasoning depth required, dynamically applying stronger vector steering when navigating intricate, multi-hop evidence spaces.

\begin{figure}
  \includegraphics[width=\columnwidth]{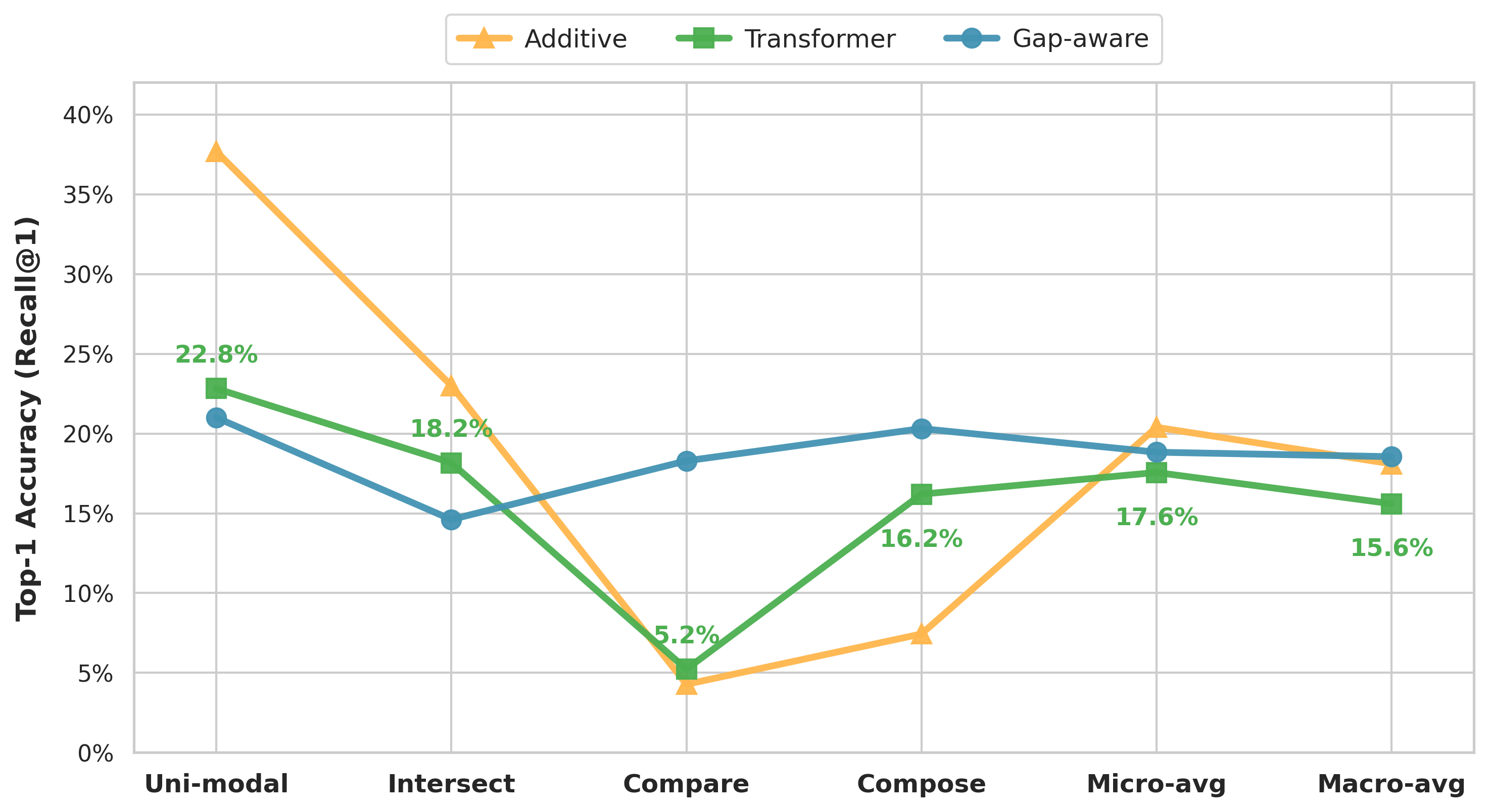}
  \caption{\textbf{Baseline performance comparison across question categories.} Comparison of Additive, Transformer-based Early Fusion, and Gap-aware (GRAIL) models across different question types (R@1 in MultimodalQA). Gap-aware maintains consistent performance on complex reasoning tasks (\textit{Compare} and \textit{Compose}) and achieves the highest overall Macro-average.}
  \label{fig:transformer}
\end{figure}

\paragraph{Comparison with Transformer-based Baseline}
To examine whether sophisticated context modeling resolves the category-specific limitations of conventional additive approaches, we evaluate a Transformer-based Early Fusion baseline \cite{vaswani-2017-attention} utilizing multi-head self-attention. Fundamentally, Transformer-based self-attention belongs to the same paradigm of conventional additive approaches, as it inherently aggregates and accumulates context vectors during feature interaction. Consequently, as shown in Figure~\ref{fig:transformer}, applying self-attention does not eliminate the sharp performance trade-offs across different question categories; like traditional additive models, the Transformer baseline exhibits sensitivity to query types, performing adequately on simpler tasks but suffering distinct performance drops on complex ones. Crucially, in reasoning-heavy multi-hop categories like Compare, the Transformer baseline remains bottlenecked by feature accumulation and semantic anchoring (5.2\% on R@1). In contrast, Gap-aware demonstrates a decisive structural advantage, maintaining robust performance where conventional attention-based aggregation falters. This confirms that explicitly isolating the missing latent gap is fundamentally more effective for cross-modality retrieval than conventional feature accumulation.

\section{Generalization on WebQA} \label{sec:app_webqa}
To verify the dataset-agnostic scalability and explore the boundary conditions of our framework, we evaluate GRAIL on the WebQA dataset (Corpus statistics detailed in Table~\ref{tab:webqa_corpus_stats} and Table~\ref{tab:webqa_task_split_stats}). Distinct from MultimodalQA which demands complex cross-modal fusion per query, WebQA comprises single-modality tasks (retrieving either multiple text snippets or multiple images) scattered across a heterogeneous collection.

While purely uni-modal retrieval is technically outside the primary scope of this work -- which targets compositional cross-modal multi-hop QA -- we include WebQA specifically to examine GRAIL's operational limits under a baseline-favored, uni-modal regime.

\begin{table}[h]
\centering
\small
\caption{Statistics of the WebQA corpus.}
\label{tab:webqa_corpus_stats}
\begin{tabular}{l|cc|c}
\hline
\textbf{Modality} & \textbf{Text} & \textbf{Image} & \textbf{Total} \\ \hline
\textbf{Count} & 666,998 & 389,740 & 1,056,738 \\ \hline
\end{tabular}
\end{table}

\begin{table}[h]
\centering
\small
\setlength{\tabcolsep}{3pt}
\caption{Statistics of the WebQA question type.}
\label{tab:webqa_task_split_stats}
\begin{tabular}{l|rrr|r}
\hline
\textbf{Question Type} & \textbf{Train} & \textbf{Val} & \textbf{Test} & \textbf{Total} \\ \hline
text & 32,827 & 4,153 & 4,169 & 41,149 \\
number & 264 & 60 & 36 & 360 \\
YesNo & 6,848 & 790 & 844 & 8,482 \\
choose & 4,590 & 608 & 516 & 5,714 \\
color & 834 & 70 & 108 & 1,012 \\
shape & 136 & 6 & 14 & 156 \\
Others & 2,514 & 316 & 322 & 3,152 \\ \hline
\textbf{Total Inst.} & 48,013 & 6,003 & 6,009 & 60,025 \\ \hline
\end{tabular}
\end{table}

\begin{table}[h]
\centering
\footnotesize
\setlength{\tabcolsep}{1.5pt}
\caption{\textbf{Distribution of gating coefficients on WebQA.} Values are reported as mean (standard deviation). Dynamic gating adapts vector steering intensity across various single-modality query categories.}
\label{tab:webqa_gate_mix_stats}
\begin{tabular}{lc}
\toprule
\textbf{Category} & \textbf{Gate} $g$ \\
\midrule
text   & $0.231\ (0.081)$ \\
number & $0.092\ (0.087)$ \\
YesNo  & $0.097\ (0.094)$ \\
choose & $0.172\ (0.119)$ \\
color  & $0.133\ (0.127)$ \\
shape  & $0.077\ (0.067)$ \\
Others & $0.089\ (0.090)$ \\
\midrule
\textbf{Overall} & ${0.197\ (0.104)}$ \\
\bottomrule
\end{tabular}
\end{table}

As shown in Table~\ref{tab:webqa_gate_mix_stats}, the learned gating parameter $g$ maintains an overall average of $0.197$ on WebQA, modulating intensity according to the reasoning demands of each query type. Furthermore, pre-alignment evaluation (Table~\ref{tab:webqa_alignment}) reconfirms that the frozen \textit{Query-evidence} topology serves as the optimal structural foundation (25.56\% for R@1, 43.03\% for R@5), mirroring our architectural findings on MultimodalQA.

\begin{table}[h]
\centering
\footnotesize
\setlength{\tabcolsep}{2pt}
\caption{\textbf{Retrieval performance of gap-aware mechanism across different alignment strategies in WebQA.} R@K metrics are reported in percentages (\%). Full empirical comparison across all cross-modal topologies under both frozen and fine-tuned training regimes. (\texttt{FT}: full fine-tuning, \texttt{Frz}: frozen backbones). Mapping individual assets directly to the query via frozen \textit{Query-evidence} serves as the optimal topological setting.}
\label{tab:webqa_alignment}
\begin{tabular}{llccccc}
\hline
\textbf{Alignment} & & \textbf{R@1} & \textbf{R@5} & \textbf{R@20} & \textbf{R@50} & \textbf{$\Delta_{\text{esc}}$} \\ \hline
None           & Frz & \underline{21.05} & \underline{41.25} & 56.66 & \underline{64.91} & $-0.165$ \\
               & FT  & 20.89 & 40.71 & 56.40 & 64.63 & $-0.265$ \\ \hline
Centroid       & Frz & 20.30 & 40.57 & 55.43 & 64.12 & $-0.276$ \\
               & FT  & 20.46 & 40.35 & 55.67 & 63.88 & $-0.265$ \\
CLS            & Frz & 7.07  & 23.47 & 38.65 & 48.74 & $-0.175$ \\
               & FT  & 7.21  & 22.84 & 37.91 & 47.96 & $-0.330$ \\ \hline
Query-evidence & Frz & \textbf{25.56} & \textbf{43.03} & \textbf{57.06} & {64.71} & \textbf{+0.038} \\
               & FT  & 20.89 & 40.71 & 56.40 & 64.63 & $-0.467$ \\
Query-set      & Frz & 15.69 & 39.51 & \underline{56.74} & \textbf{65.31} & \underline{$-0.126$} \\
               & FT  & \underline{21.05} & 40.85 & 56.12 & 64.47 & $-0.454$ \\ \hline
\end{tabular}
\end{table}

When retrieving within the exact same modality, captured context and target evidence often reside in densely clustered, near-collinear subspaces. Under such conditions, overly aggressive subtractive steering risks erasing shared intra-modality features--a phenomenon we term \textit{over-subtraction}.

As shown in Table~\ref{tab:webqa_category_eval}, this boundary behavior creates a performance trade-off. As previously observed in the MultimodalQA dataset experiments, the \textit{Additive} model shows strength when retrieving evidence within single modality. Because individual queries in WebQA focus on finding multiple assets from the =same modality (either only text snippets or only images), this dataset is also likely to favor the \textit{Additive} baseline. This allows Additive to perform better in minor attribute tasks such as \textit{Number}, \textit{YesNo}, and \textit{Shape}.

However, GRAIL shows a massive reversal when given enough training data to learn this non-cross-modal environment. In data-rich categories (\textit{Text}, $68.37\%$ of train split), where sufficient optimization data exists, GRAIL successfully calibrates its subtractive intensity to avoid over-subtraction, achieving a massive $+25.98$pp boost in R@1 ($37.95\%$ vs. $11.97\%$). Conversely, in data-sparse categories like \textit{Number} (0.28\%), with only 136 training instances, the gating module struggles to fine-tune its projection bounds, leaving them reliant on Additive's simple token accumulation.

Crucially, despite raw recall fluctuations in sparse categories, the semantic escape delta($\Delta_{\text{esc}}$) metric highlights the fundamental fragility of additive fusion. While the \textit{Additive} baseline yields negative margins across every single category (ranging from $-0.134$ to $-0.229$), GRAIL maintains strictly positive $\Delta_{\text{esc}}$ values throughout ($+0.012$ to $+0.043$), proving its superior precision in isolating missing information.

\begin{figure}[t]
  \includegraphics[width=\columnwidth]{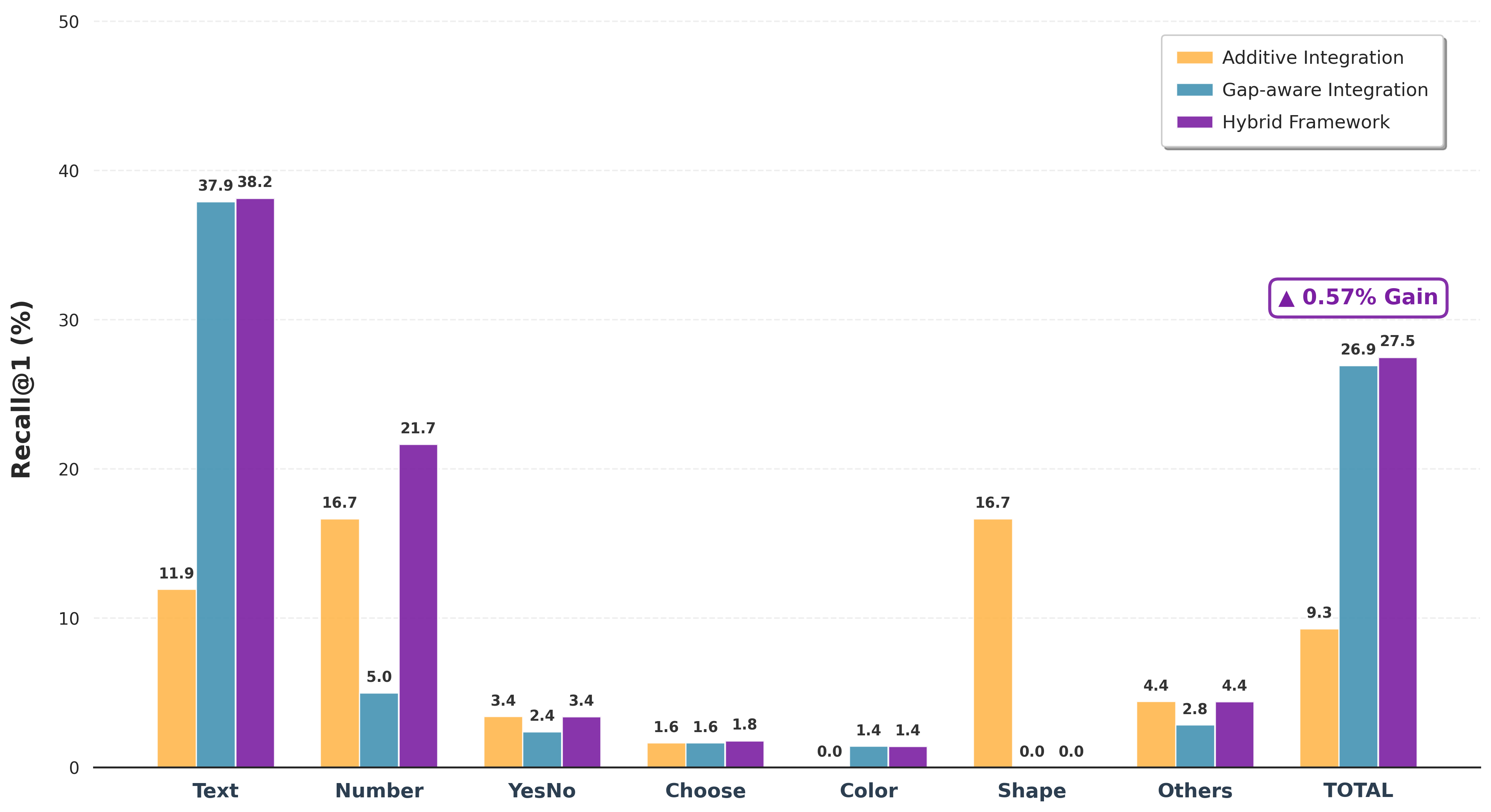}
  \caption{\textbf{Hybrid framework leveraging complementary specialists in WebQA.} By dynamically routing each query to the optimal specialist, our Hybrid framework achieves a "best-of-both-worlds" performance. (+195.7\% micro avg. gain over Additive)}
  \label{fig:webqa_hybrid}
\end{figure}

This empirical complementarity confirms that no single vector operator universally dominates all retrieval conditions. Consequently, we introduce a Hybrid framework guided by a lightweight question-type selector. As illustrated in Figure~\ref{fig:webqa_hybrid}, this hybrid architecture routes data-rich reasoning queries to GRAIL while delegating localized uni-modal tasks to the Additive specialist, achieving a unified ``best-of-both-worlds'' performance ($+18.2$pp micro-average gain over Additive).

\begin{table}
\centering
\footnotesize
\setlength{\tabcolsep}{2pt}
\caption{\textbf{Retrieval performance across question types in WebQA.} Comparison between the Additive baseline and GRAIL subject to q\_type.}
\label{tab:webqa_category_eval}
\begin{tabular}{llccccc}
\hline
\textbf{Type} & \textbf{Method} & \textbf{R@1} & \textbf{R@5} & \textbf{R@10} & \textbf{R@20} & \textbf{$\Delta_{\text{esc}}$} \\ \hline
\multirow{2}{*}{Text} & Additive & 11.97 & 49.87 & 63.64 & 74.14 & $-0.228$ \\
 & GRAIL & \textbf{37.95} & \textbf{60.77} & \textbf{69.30} & \textbf{75.85} & \textbf{$+$0.043} \\ \hline
\multirow{2}{*}{Number} & Additive & \textbf{16.67} & \textbf{30.00} & \textbf{38.33} & \textbf{40.00} & $-0.141$ \\
 & GRAIL & 5.00 & 11.67 & 16.67 & 16.67 & \textbf{$+$0.014} \\ \hline
\multirow{2}{*}{YesNo} & Additive & \textbf{3.42} & \textbf{12.78} & \textbf{18.10} & \textbf{25.57} & $-0.205$ \\
 & GRAIL & 2.41 & 6.71 & 10.51 & 16.58 & \textbf{$+$0.014} \\ \hline
\multirow{2}{*}{Choose} & Additive & 1.64 & 4.77 & 7.57 & 10.86 & $-0.229$ \\
 & GRAIL & 1.64 & \textbf{5.10} & \textbf{8.06} & \textbf{12.83} & \textbf{$+$0.032} \\ \hline
\multirow{2}{*}{Color} & Additive & 0.00 & \textbf{10.00} & \textbf{21.43} & \textbf{24.29} & $-0.210$ \\
 & GRAIL & \textbf{1.43} & 8.57 & 8.57 & 17.14 & \textbf{$+$0.020} \\ \hline
\multirow{2}{*}{Shape} & Additive & \textbf{16.67} & \textbf{66.67} & \textbf{66.67} & \textbf{83.33} & $-0.134$ \\
 & GRAIL & 0.00 & 16.67 & 33.33 & 66.67 & \textbf{$+$0.020} \\ \hline
\multirow{2}{*}{Others} & Additive & \textbf{4.43} & \textbf{12.97} & \textbf{18.67} & \textbf{23.73} & $-0.203$ \\
 & GRAIL & 2.85 & 8.54 & 12.03 & 18.67 & \textbf{$+$0.012} \\ \hline
\end{tabular}
\end{table}

Finally, Sequential Pool Construction evaluations on WebQA (Table~\ref{tab:webqa_iterative}) demonstrate GRAIL's structural noise-resilience over multi-step execution. In WebQA, where queries typically target roughly two evidence pieces and strong baselines easily capture the first gold item, the \textit{Additive} model completely collapses under cumulative noise ($-1.50$pp in \texttt{5+5} and $-2.53$pp in \texttt{3+7}). In contrast, GRAIL's subtractive operator explicitly cancels out already-retrieved context, shielding the search vector from catastrophic drift and yielding stable positive gains under fine-grained partitioning schedules (+0.27pp Set-Recall and +0.46pp NRM under \texttt{4+3+3}).

\begin{table}
\centering
\small
\caption{\textbf{Granular efficiency under various step configurations ($K = 10$) in WebQA.} Relative improvements are measured related to Query Only baseline. GRAIL maintains robust, positive margins across granular partition schedules, whereas the Additive framework suffers severe performance degradation.}
\label{tab:webqa_iterative}
\begin{tabular}{lccc}
\hline
\textbf{Steps} & \textbf{Set-Rec@10} & \textbf{Jump} & \textbf{NRM} \\ \hline
\textit{\textbf{Query Only}} &                  &                     &              \\ \hline
\texttt{10} & 51.06\%& --  & --\\ \hline
\textit{\textbf{Additive}} &                  &                     &              \\ \hline
\texttt{5+5} & $-1.50$pp & \underline{25.3} & $-2.06$pp    \\
\texttt{3+7} & $-2.53$pp & \textbf{31.9} & $-3.61$pp \\ \hline
\textit{\textbf{GRAIL (Ours)}} &              &                     &              \\ \hline
\texttt{8+1+1}& \underline{$+$0.25pp}& 15.4 & \underline{$+$0.59pp}      \\
\texttt{8+2}            & $+$0.23pp             & 16.4               & \textbf{$+$0.69pp} \\
\texttt{5+5}            & $-0.22$pp           & 16.1                & $-0.20$pp    \\
\texttt{3+7}            & $-0.53$pp           & 15.5              & $+$0.00pp       \\
\texttt{4+3+3}          & \textbf{$+$0.27pp}    & 16.4               & $+$0.46pp      \\
\texttt{3+2+3+2}        & $+$0.10pp             & 16.2               & $+$0.45pp      \\
\texttt{2*5}           & $-0.02$pp           & {16.7}                & $-0.13$pp    \\
\texttt{1*10}           & $+$0.05pp             & {17.0}                & $-0.11$pp    \\ \hline
\end{tabular}
\end{table}

\section{Inference Cost Comparison}
\label{sec:app_cost}

\begin{table*}
\centering
\footnotesize
\setlength{\tabcolsep}{2.5pt}
\caption{\textbf{Inference efficiency and operational cost comparison.} All configurations utilize precomputed asset embeddings, requiring zero evidence encoder passes during inference. 
GRAIL completely eliminates the autoregressive LLM inference loop required by explicit query generation baselines, achieving millisecond-level latency and zero API cost while maintaining robustness to retrieval noise.}
\label{tab:cost_efficiency}
\vskip 0.1in
\begin{tabular}{lcccccc}
\toprule
\textbf{Model/ Cost Dimension} & \textbf{Query Only} & \textbf{Additive} & \textbf{GRAIL (ours)} & \textbf{No-Ctx} & \textbf{IRCoT} & \textbf{Ans-Aug} \\
\midrule
LLM API Calls (per Q) & 0 & 0 & 0 & 1 & 1 & 1 \\
Question Encoder Passes & 1 & 1 & 1 & 1 & 1 & 1 \\
Active Parameter Scale & $\sim$335M & $\sim$335M & $\sim$335M+1.3M & \multicolumn{3}{c}{--- $\sim$335M + LLM (7B--100B) ---} \\
Trainable Parameters & 3.68M & 3.68M & 5.00M (3.68M + 1.32M) & \multicolumn{3}{c}{--- 3.68M (Shared Alignment Base)$^\ddagger$ ---} \\
\midrule
Avg. Input Tokens & 0 & 0 & 0 & $\approx$ 100 & $\approx$ 776 & $\approx$ 776 \\
Avg. Output Tokens & 0 & 0 & 0 & $\approx$ 15 & 35.6 & 42.6 \\
\midrule
Avg. Latency (per Q) & 2--5 ms & 2--5 ms & 3--6 ms & 0.6--1.0 s & 1.1--1.6 s & 1.2--1.8 s \\
Relative Throughput & $\sim$400$\times$ & $\sim$400$\times$ & $\sim$350$\times$ & 1.8$\times$ & 1.0$\times$ (Base) & 0.9$\times$ \\
API Cost (per 1K Qs)$^\dagger$ & \$0.00 & \$0.00 & \$0.00 & \$0.024 / \$0.40 & \$0.138 / \$2.30 & \$0.142 / \$2.37 \\
\midrule
Vulnerability to Noise & N/A & High & Low & N/A & Moderate & Moderate \\
\bottomrule
\end{tabular}
\vskip 0.05in
\raggedright \footnotesize $^\dagger$ Costs are formatted as \texttt{GPT-4o-mini} / \texttt{GPT-4o} tiers.
\raggedright \footnotesize $^\ddagger$ LLM-based models utilize the frozen 3.68M parameters from our optimized alignment phase to perform the initial dense asset retrieval.
\end{table*}

Based on the efficiency profiles in Table~\ref{tab:cost_efficiency}, we evaluate the computational and financial viability of GRAIL against explicit query regeneration via LLMs.

\paragraph{Latency and Throughput Disparity} 
Explicit LLM query regeneration introduces severe inference bottlenecks due to its autoregressive generation. As reported in Table~\ref{tab:cost_efficiency}, \textit{IRCoT} and \textit{Answer-Augmented Retrieval} require an average of 1.1--1.6 and 1.2--1.8 seconds per question, heavily burdened by processing $\approx$776 input tokens and generating up to 42.6 output tokens per step. 
In contrast, GRAIL completely bypasses the LLM inference loop during retrieval. By executing context-subtractive query steering via a lightweight gating mechanism, our framework achieves a near-instantaneous latency of just 3--6 ms per question. This translates to a \textbf{$\sim$350$\times$ throughput advantage}, enabling real-time, large-scale multimodal retrieval.

\paragraph{Financial and API Cost Analysis}
The operational expenses further highlight the scalability of GRAIL. LLM-based query regeneration incurs substantial commercial API costs that scale linearly with token volume. For instance, evaluating 1K questions under \textit{IRCoT} costs up to \$2.30 via premium tiers like \texttt{GPT-4o}, and remains a noticeable overhead (\$0.138) even with lightweight alternatives like \texttt{GPT-4o-mini}. 
Conversely, GRAIL operates at a strict \textbf{\$0.00 API cost} by relying solely on local bi-encoder operations over precomputed embeddings. By replacing explicit text-based query generation with vector subtraction, our method eliminates financial dependencies on external commercial APIs, providing a fully self-contained deployment framework.

\end{document}